\documentclass[twocolumn]{aastex63}

\font\smcap=cmcsc10

\newcommand{\kms}{\,km~s$^{-1}$}
\newcommand{\hi}{H\,{\smcap i}}

\received{November 12, 2021}
\revised{January 28, 2021}
\accepted{February 7, 2022}
\submitjournal{ApJ}

\shorttitle{M33 Disk Dynamics}
\shortauthors{Quirk et al.}
\graphicspath{{./}{figures/}}

\begin{document}

\title{The Triangulum Extended (TREX) Survey: The Stellar Disk Dynamics of M33 as a Function of Stellar Age}

\correspondingauthor{Amanda C. N. Quirk}
\email{acquirk@ucsc.edu}

\author[0000-0001-8481-2660]{Amanda C. N. Quirk}
\affiliation{UCO/Lick Observatory, Department of Astronomy \& Astrophysics, University of California Santa Cruz, 
 1156 High Street, 
 Santa Cruz, California 95064, USA}

\author{Puragra Guhathakurta}
\affiliation{UCO/Lick Observatory, Department of Astronomy \& Astrophysics, University of California Santa Cruz, 
 1156 High Street, 
 Santa Cruz, California 95064, USA}
 
 \author[0000-0003-0394-8377]{Karoline M. Gilbert}
\affiliation{Space Telescope Science Institute, 3700 San Martin Dr., Baltimore, MD 21218, USA}
\affiliation{Department of Physics \& Astronomy, Bloomberg Center for Physics and Astronomy, Johns Hopkins University, 3400 N. Charles Street, Baltimore, MD 21218}

\author[0000-0002-3834-7937]{Laurent Chemin}
\affiliation{Centro de Astronom\'ia, Universidad de Antofagasta \\
Avda. U. de Antofagasta 02800, Antofagasta, Chile}

\author[0000-0002-1264-2006]{Julianne J. Dalcanton}
\affiliation{Department of Astronomy, University of Washington, Box 351580, U.W., Seattle, WA 98195-1580, USA}
\affiliation{Center for Computational Astrophysics, Flatiron Institute, 162 Fifth Avenue, New York, NY 10010, USA}

\author[0000-0002-7502-0597]{Benjamin F. Williams}
\affiliation{Department of Astronomy, University of Washington, Box 351580, U.W., Seattle, WA 98195-1580, USA}

\author[0000-0003-0248-5470]{Anil Seth}
\affiliation{
Dept. of Physics Astronomy, 
University of Utah, 
115 South 1400 East, 
Salt Lake City, UT 84112, USA}

\author{Ekta Patel}
\affiliation{University of California, Berkeley, 501 Campbell Hall, Berkeley, CA, 94720}
\affiliation{Miller Institute for Basic Research in Science, 468 Donner Lab, Berkeley, CA 94720}

\author{Justin T. Fung}
\affiliation{The Harker School, 500 Saratoga Ave., San Jose, CA 95129, USA}
\author{Pujita Tangirala}
 \affiliation{Saint Francis High School, 1885 Miramonte Avenue
Mountain View, California 94040, USA}
 \author{Ibrahim Yusufali}
 \affiliation{Duke University, 2080 Duke University Road, Durham, North Carolina 27708, USA}



\begin{abstract}
Triangulum (M33) is a low mass, relatively undisturbed spiral galaxy that offers a new regime in which to test models of dynamical heating. In spite of its proximity, M33’s dynamical heating history has not yet been well constrained. In this work, we present the TREX Survey, the largest stellar spectroscopic survey across the disk of M33. We present the stellar disk kinematics as a function of age to study the past and ongoing dynamical heating of M33. We measure line-of-sight velocities for $\sim 4500$ disk stars. Using a subset, we divide the stars into broad age bins using Hubble Space Telescope and Canada-France-Hawaii Telescope photometric catalogs: massive main sequence stars and helium burning stars ($\sim 80$ Myr), intermediate mass asymptotic branch stars ($\sim 1$ Gyr), and low mass red giant branch stars ($\sim 4$ Gyr). We compare the stellar disk dynamics to that of the gas using existing \hi, CO, and H$\alpha$ kinematics. We find that the disk of M33 has relatively low velocity dispersion ($\sim 16$ \kms), and unlike in the Milky Way and Andromeda galaxies, there is no strong trend in velocity dispersion as a function of stellar age. The youngest disk stars are as dynamically hot as the oldest disk stars and are dynamically hotter than predicted by most M33-like low-mass simulated analogs in Illustris. The velocity dispersion of the young stars is highly structured, with the large velocity dispersion fairly localized. The cause of this high velocity dispersion is not evident from the observations and simulated analogs presented here.

\end{abstract}


\section{Introduction} \label{sec:intro}
The Triangulum Galaxy (M33) is one of two dwarf spirals in the Local Group \citep{Gaia_Co2021}. With a stellar mass of roughly $4.8 \times 10^{9}M_{\odot}$ \citep{corbelli2014}, it is ten times less massive than the Milky Way (MW) and Andromeda (M31) and twice as massive as the Large Magellanic Cloud (LMC). It has a high star formation rate (SFR), $\sim0.7M_{\odot} \rm yr^{-1}$ \citep{Blitz_2006}, characteristic of later type spirals. Furthermore, it is a relatively isolated galaxy, making its disk more pristine. As high redshift galaxy observations push down to lower stellar masses, M33 becomes an important local comparison point. 

\par While M33 is sufficiently close to be studied in detail, much remains unknown. For example, there is debate as to its interaction history with M31. Some of M33's properties could be explained by an interaction; such as its highly warped \hi\ disk that extends far beyond the stellar disk, out to $22\rm\ kpc$ \citep{braun2004, putman2009, semczuk2018}, the apparent disturbances to the outskirts of M33's stellar disk \citep{mcconnachie09, McConnachie_2010}, and the mutual burst of star formation $2-4$ Gyr ago for both M31 and M33 \citep[e.g.,][]{putman2009, mcconnachie09, lewis2015, Ferguson2016, semczuk2018, Williams2017}. However, new studies using the proper motion of both galaxies from Gaia DR2, the Hubble Space Telescope (HST) and the Very Long Baseline Array (VLBA), and cosmological simulations support the notion that M33 is on its first infall \citep{patel2017a, patel2017b, vandermarel18} and has had no significant interaction with M31. 

\par Furthermore, M33 is more complex then expected for a low luminosity galaxy, which are thought to be dominated by a single component, although the highest mass dwarfs can have a thick disk in addition to a stellar halo \citep{Roychowdhury2013, van_der_Wel_2014, Patra2020, Kado_Fong_2020, kado-fong2021}. However, M33, is proving to be more complex, with a newly confirmed centrally-concentrated halo \citep{Gilbert2021} and bar \citep{Williams2021, smercina2022, lazzarini2022} and the mysterious warps described above \citep{braun2004, putman2009, semczuk2018}.

\par Resolved stellar spectroscopy is a valuable tool to unlock the history of M33, especially when information from spectroscopic observations is examined as a function of stellar age. This is {\it{only}} possible in the Local Group, in which the distances to galaxies allow us to view their entire disk while resolving individual stars. For example, stellar line-of-sight velocity dispersion as a function of stellar age can point to gradual heating and heating via one event, as seen in M31, where the high velocity dispersion increases monotonically with stellar age, suggesting a constant and violent merger history \citep{Dorman2015}. Comparing stellar kinematics to gas kinematics can also give insight into a galaxy's dynamical heating history, as was demonstrated in M31 \citep{Quirk2019}, as the asymmetric drift of stellar populations is correlated with their velocity dispersion in the plane of the disk. Furthermore, comparing this observed asymmetric drift to that in simulations suggests that M31 had a relatively recent 4:1 merger \citep{Quirk2020}. M33's distance of  $859 \rm\ kpc$ \citep[]{degrijs2014} is comparable to that of M31 \citep{mcconnachie09}, so we can measure spectroscopy of individually resolved stars in M33. These techniques may also be capable of constraining the merger history of M33.

\par We expect M33 to be a particularly interesting target for these studies. Unlike M31's obvious remnants from its violent history, M33 is morphologically less disturbed. It has however a much higher SFR \citep{Blitz_2006} and lower disk mass surface density \citep{corbelli2014}, placing its disk in a very different regime than M31's. M33 is also the prime environment to observe the effects of internal heating (i.e., bursts from star formation, perturbations from giant molecular clouds and the bar, and density waves from spiral arms) because of its low mass and low inclination \citep[$\sim54 \degr$]{warner1973}. Stellar feedback can cause powerful inflows and outflows of gas and bursts of star formation, which can result in drastic radial migration in low mass galaxies and can even lead to the creation of a stellar halo \citep[e.g.,][]{stinson2009, maxwell2012, el-badry2016}. This internal feedback could have drastically changed the stellar kinematics of M33 since its birth. Stellar disks are fragile \citep{toth1992}, and while more massive disks are likely able to survive major merger events \citep[e.g.,][]{DSouza2018, Hammer2018} and low mass disks are believed to have to survive many minor events \citep{helmi2012}, low mass disks, like that of M33, are unlikely to remain intact after major merging events. This notion paired with the fact that M33 is relatively isolated ($\sim 230\rm\ kpc$ from M31), and about half of its stellar mass comes from stars that are $\sim 10$ Gyr \citep{Williams_2009}, means that the disk of M33 is fairly pristine and therefore can give us insight into the evolution of isolated high redshift galaxies. 

\par {\it In this paper, we present the TRiangulum EXtended (TREX) Survey of $\sim 7000$ targets, making it the largest stellar spectroscopic survey of M33.} The survey spans across the entire disk of M33, out to a deprojected radius of $\sim 11$ kpc. It is the first dataset that consists of individually resolved stars that extends across the entire inner and into the outer disk. {\it With this dataset, we examine the kinematics of stars in the disk of M33 as a function of stellar age to measure the dynamical heating of the evolving disk.} This analysis, which uses a subset of the total sample, is the first study of disk kinematics as a function of stellar age in M33 using only individual stars and is overall the third of its kind (after the MW and M31). The robust dataset presented here has already been used to confirm the existence of a dynamically hot component in M33 \citep{Gilbert2021}.
\par This paper is organized as follows. In Section \ref{sec:Data} we present our new spectroscopic dataset and archival datasets used in this study. Section \ref{sec:ages} describes the separation of stars into broad age bins and the removal of possible halo stars, and Section \ref{sec:velocities} shows the calculation of local velocity dispersion, rotation velocity, and asymmetric drift. Section \ref{sec:illustris} highlights a comparison of observed kinematics to the kinematics seen in M33-like simulated galaxies, and Section \ref{sec:LG} compares them to the kinematics of M31 and the MW Solar Neighborhood. We summarize the main points of this work in Section \ref{sec:summary}, and in Section \ref{sec:rare}, we show the kinematics of rare spectral types.

\section{Data} \label{sec:Data}
Our study made use of large catalogs of stellar photometry and spectroscopy, as well as imaging data of the M33 gas content. Below we describe the photometric, stellar spectroscopic, and gas imaging catalogs.

\subsection{Stellar Data} 
 We started with large libraries of resolved stellar data, including space-based and ground-based photometry. These, in turn, allowed us to obtain our stellar spectroscopic dataset. We selected targets from a wide range of masses and evolutionary stages, including massive main sequence (MS), massive helium burning (HeB), intermediate mass asymptotic giant branch (AGB), and low mass red giant branch (RGB) stars. The use of these different stellar types is described later in Section \ref{sec:ages}. These different stages were not prioritized equally over the four observing epochs, see description below and Table \ref{tab:masks_info}. The broad evolutionary stage of a star comes from color magnitude diagrams (CMD) of the photometry catalogs described in Section \ref{sec:phot}. We describe this process in detail below. 
 
\subsubsection{Photometry} \label{sec:phot}
Our strategy for target selection relies on selecting bright isolated stars from resolved stellar photometry. The high precision of this photometry allows us to target stars in the crowded central regions of M33 with confidence that we were observing isolated stars instead of blended light from multiple sources. 
\par Over the four years of spectroscopic observations, our stellar selection varied in response to the available photometry and evolving scientific opportunities, as stated in Table \ref{tab:masks_info}. We relied on a mix of photometry from the Hubble Space Telescope (HST) and the Canada-France-Hawaii Telescope (CFHT). Targets observed in 2016 were selected using archival HST data with broad bands F475W + F814W, or F606W + F814W or  where there was a gap in HST coverage, using data from MegaCam on CFHT with $i-$ and $g-$ bands. The HST fields were observed with the Advanced Camera for Surveys (ACS) and the reduction is described in \cite{Williams_2009, Williams2014}. Each of the 2016 masks overlapped with multiple of these ACS fields. The CFHT/MegaCam data was reduced using the MegaPipe reduction pipeline \citep{gwyn2008}. The primary targets for these masks were RGB stars, but also included some AGB and red HeB stars and a small number of MS and BHeB stars.
\par The 2018 and 2019 slitmasks had targets selected from HST photometry from the Panchromatic Hubble Andromeda Treasury: Triangulum Extended Region \cite[PHATTER]{Williams2021} and CFHT/MegaCam (same as described above). The PHATTER survey observed stars in the Andromeda and Triangulum galaxies with six filter photometry using ACS and WFC3: F275W, F336W, F475W, F814W, F110W, and F160W \citep{Dalcanton2012, Williams2021}. The photometric catalogs are described in \cite{Williams2014, Williams2021}. 
\par The availability of six filter photometry allows us to more precisely divide stars into broad age bins, as described later in Section \ref{sec:ages}. With the six filter photometry, we were able to target a range of stellar evolutionary stages for the 2018 masks: MS, HeB, AGB, and RGB stars. To sample a broad range of stellar ages, we preferentially targeted rarer stars, including a large number of bright HeB stars. These stars were identified from PHATTER CMDs. CFHT/MegaCam data was used to fill in the outer parts of the DEIMOS slitmasks that extended beyond the HST PHATTER footprint, into the low density outer disk where HST resolution is less needed. For the 2019 masks, RGB stars were the primary targets. These stars were identified from PHATTER CMDs if in the PHATTER range. If they came from the CFHT/MegaCam MegaPipe reduction, we prioritized stars with $g - i > 0.5$ and $21 < i < 22$.
\par The 2020 slitmasks were positioned to probe the outer disk, and they are beyond the PHATTER footprint and any other continuous HST coverage. For these outer slitmasks, we used the catalog from the Pan-Andromeda Archaeological Survey (PAndAS) \citep{McConnachie2018}. PAndAS used CFHT/Megacam to observe $>$400 square degrees of sky centered on Andromeda and Triangulum with $i-$ and $g-$ bands. The observations and data reduction are described in \cite{McConnachie2018}. Only objects that were flagged by the PAndAS team to most likely be stars were included in our target list. We prioritized placing RGB stars on these masks ($g - i > 0.5$ and $20.5 < i < 22.5$).

\par To avoid blends, especially in the crowded central regions, we applied an isolation criterion $I_{neighbor} < I_{target} - (\frac{d}{0.8\arcsec})^{2} +3$ to exclude stars with neighbors that are too close and/or too bright and therefore might contaminate the target's light during spectroscopic observation with DEIMOS\citep{Dorman2012}. We applied this criterion to all of the photometry catalogs used, although it was most critical for the crowded regions targeted using PHATTER photometry. If a target candidate has a single neighbor that fulfills this criterion, it is excluded from the slitmask. 

\par Even with this criterion, it is possible to have multiple objects in a given slit. The majority of these serendipitous observations have good quality spectra and well-measured velocities that did not interfere with the main target's spectrum. However, since we do not have easily paired photometry for these objects, we do not include them in this particular study, although they will eventually be incorporated. The total number of targets is 7684, with 2381 from 2016, 2457 from 2018, 906 from 2019, and 1940 from 2020. Adding in the serendipitous targets would increase the sample by $\sim 27 \%$.

\subsubsection{Keck II DEIMOS Spectroscopy}\label{sec:spec}

The spectroscopic data comes from four epochs of observing. All observations were taken with the DEIMOS Spectrograph \citep{faber2003} on the Keck II 10 meter telescope. The program uses thirty-six DEIMOS slitmasks and two different grating setups-- one to target a wide range of stellar evolutionary phases and one to target older, redder stars. The first ten slitmasks were observed in 2016 using the 600 line mm$^{-1}$ grating, which has a resolution of R$\sim 2000$, and a central wavelength and wavelength range of $7200$\AA and $\lambda\sim 4600$--9800\ \AA, respectively. This setting allows us to target a wide range of spectral types. In 2018, we observed eleven slitmasks across the central disk using the same configuration. In 2019, we obtained four additional slitmasks of spectroscopic data. The first used the 600 line mm$^{-1}$ grating configuration, and the remaining three were observed using the 1200 line mm$^{-1}$ grating (R$\sim 6000$)\ to account for additional moon light and to target older stars. With this grating, we used a central wavelength of 7800\AA\ and a wavelength range of $\lambda\sim 6300$--9800\ \AA, focusing on the redder part of the spectrum where RGB stars emit significant flux. In the Fall of 2020, we observed the last eleven slitmasks in the outer disk with the 1200 line mm$^{-1}$ configuration. 

\par
\begin{figure}[bp!]
    \centering
    \includegraphics[width=\columnwidth]{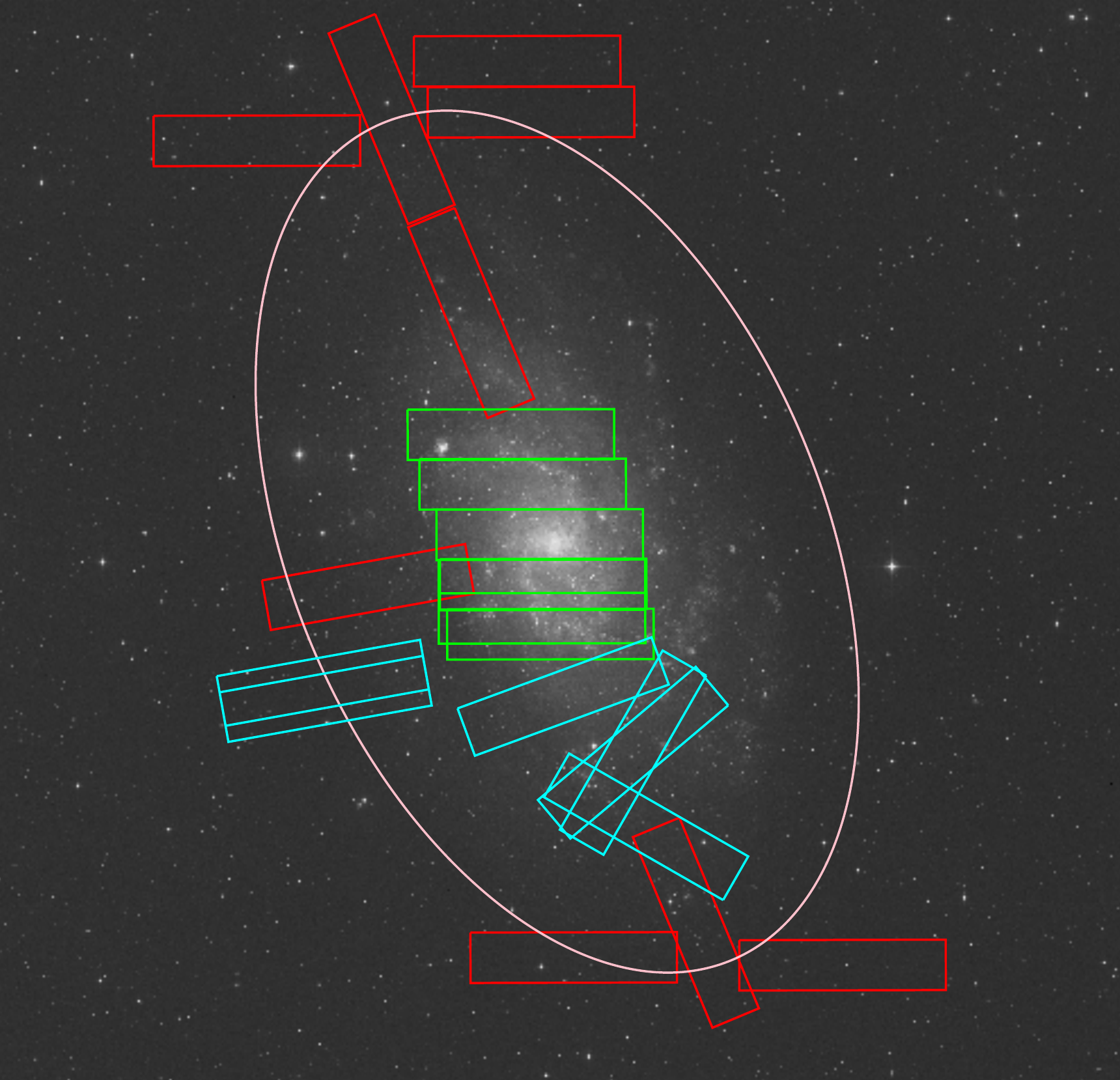}
    \caption{Layout and orientation of the thirty six DEIMOS slitmasks across the disk of M33 for the TREX Survey. Each rectangle on the image represents the rough shape and size of a DEIMOS slitmask (approximately 16\arcmin by 4\arcmin). The cyan bricks were observed in 2016, the green in 2018 and 2019, and the red in 2020. Because of the high density of stars in the central regions, we targeted different stars at the same slitmask position in 2018 and 2019. The 2019 slitmasks placements are the top two green slitmasks. The ellipse represents the approximate location of a break in the exponential surface density profile of the disk at $\sim 36$\arcmin\ \citep{ferguson2007, barker2011}. The background image of M33 is from the Digital Sky Survey (DSS). In this orientation, north is up, and east is to the left.}
    \label{fig:m33_masks}
\end{figure}

 The layout of the thirty six slitmasks can be seen in Figure \ref{fig:m33_masks}. Because of the high density of stars in the inner regions, we were able to have slitmasks targeting different stars at the same slitmask location and orientation. Some of the targets on different slitmasks were repeated to get higher signal for faint targets and to help calibrate velocity measurement errors. The positions of the 2019 slitmasks are the same as the two most northern 2018 slitmasks. Each slitmask was observed for approximately 2 hours. Table \ref{tab:masks_info} lists the names, positions, orientations, exposure times, numbers of stars observed, years observed, gratings used, photometry sources for target selection, and the main targets for each DEIMOS slitmask. 

\begin{deluxetable*}{c|c|c|c|c|c|c|c|c}
    \rotate
    \centering
    \tablehead{Name & Center (J200) & Mask PA ($^\circ$)  & Exposure Time (min) & N Targets & Year & Grating (l mm$^{-1}$) & Photometry Source & Primary Targets}
    \startdata
    M33D2A & 1:33:47.95   30:27:18.8 & 110 & 76 & 246 & 2016 & 600 & Archival HST + CFHT & RGB  \\
    M33D2B & 1:33:47.95   30:27:18.8 & 110 & 88 & 244 & 2016 & 600 & Archival HST + CFHT & RGB \\
    M33D3A & 1:33:22.21   30:22:50.3 & 310 & 72 & 250 & 2016 & 600 & Archival HST + CFHT & RGB \\
    M33D3B & 1:33:22.21   30:22:50.3 & 330 & 70 & 252 & 2016 & 600 & Archival HST + CFHT & RGB \\
    M33D3D & 1:33:22.21   30:22:50.3 & 310 & 70 & 244 & 2016 & 600 & Archival HST + CFHT & RGB \\
    M33D4A & 1:33:17.50   30:16:56.9 & 240 & 70 & 239 & 2016 & 600 & Archival HST + CFHT & RGB  \\
    M33D4B & 1:33:17.50   30:16:56.9 & 240 & 84 & 227 & 2016 & 600 & Archival HST + CFHT & RGB  \\
    M33D4C & 1:33:17.50   30:16:56.9 & 240 & 76 & 233 & 2016 & 600 & Archival HST + CFHT & RGB  \\
    M33MA1 & 1:35:16.71   30:28:25.4 & 100 & 76 & 225 & 2016 & 600 & Archival HST + CFHT & RGB  \\
    M33MA2 & 1:35:15.66   30:27:08.6 & 100 & 67 & 221 & 2016 & 600 &  Archival HST + CFHT & RGB  \\
    A1M33P & 1:33:52.61   30:32:15.9 & 90 & 112 & 234 & 2018 & 600 & PHATTER + CFHT & range   \\
    A2M33P & 1:33:52.60   30:32:15.9 & 90 & 112 & 245 & 2018 & 600 & PHATTER + CFHT & range  \\
    B1M33P & 1:33:55.69   30:36:10.8 & 90 & 149.2* & 226 & 2018 & 600 & PHATTER + CFHT & range  \\
    B2M33P & 1:33:55.23   30:36:14.4 & 90 & 81.5 & 209 & 2018 & 600 & PHATTER + CFHT & range  \\
    C1M33P & 1:33:56.46   30:40:10.4 & 90 & 120 & 208 & 2018 & 600 & PHATTER + CFHT & range \\
    C2M33P & 1:33:56.46   30:40:10.4 & 90 & 110 & 200 & 2018 & 600 & PHATTER + CFHT & range \\
    D1M33P & 1:34:02.73   30:44:11.0 & 90 & 132.5 & 220 & 2018 & 600 & PHATTER + CFHT & range \\
    D2M33P & 1:34:02.73   30:44:11.0 & 90 & 110 & 213 & 2018 & 600 & PHATTER + CFHT & range \\
    E1M33P & 1:34:07.06   30:48:07.2 & 90 & 120 & 228 & 2018  & 600 & PHATTER + CFHT & range\\
    E2M33P & 1:34:07.06   30:48:07.2 & 90 & 120 & 224 & 2018  & 600 & PHATTER + CFHT & range\\
    K1M33P & 1:33:55.69   30:33:31.8 & 90 & 117.7 & 250 & 2018 & 600 & PHATTER + CFHT & range \\
    D1M33R & 1:34:02.73    30:44:11.0 & 90 & 100 & 240 & 2019 & 1200G & PHATTER + CFHT & RGB  \\
    D2M33R & 1:34:02.73   30:44:11.0 & 90 & 193* & 226 & 2019 & 1200G & PHATTER + CFHT & RGB  \\
    E1M33R19 & 1:34:07.06 30:48:07.2 & 90 & 62.5 & 201 & 2019 & 600 & PHATTER + CFHT & RGB  \\
    E2M33R & 1:34:07.06    30:48:07.2 & 90 & 100 & 239 & 2019 & 1200G & PHATTER + CFHT & RGB  \\
    pTE1 & 1:35:00.02    30:36:01.6 & 100 & 108 & 200 & 2020 & 1200G & PAndAS & RGB  \\
    pTN1a & 1:34:51.22   31:13:10.7 & 100 & 120 & 194 & 2020 & 1200G & PAndAS & RGB  \\
    pTN1b & 1:34:51.22  31:13:10.7 & 22.5 & 120 & 174 & 2020 & 1200G & PAndAS & RGB  \\
    pTN2a & 1:34:21.72 30:57:46.6 & 22.5 & 180* & 218 & 2020 & 1200G & PAndAS & RGB  \\
    pTN2b & 1:34:21.72  30:57:46.6 & 22.5 & 108 & 214 & 2020 & 1200G & PAndAS & RGB  \\
    pTN3 & 1:34:04.46     31:17:45.8 & 90 & 120 & 143 & 2020 & 1200G & PAndAS & RGB  \\
    pTN4 & 1:33:59.33   31:13:43.0 & 90 & 129.2 & 166 & 2020 & 1200G & PAndAS & RGB  \\
    pTN5 & 1:35:41.37     31:11:27.2 & 90 & 108 & 147 & 2020 & 1200G & PAndAS & RGB  \\
    pTS1 & 1:32:59.31   30:09:19.0 & 22.5 & 108 & 183 & 2020 & 1200G & PAndAS & RGB  \\
    pTS2 & 1:32:05.42      30:05:53.2 & 90 & 71 & 137 & 2020 & 1200G & PAndAS & RGB  \\
    pTS3 & 1:33:44.32     30:06:35.3 & 90 & 115 & 164 & 2020 & 1200G & PAndAS & RGB  \\
    \enddata
    \caption{Information for the 36 DEIMOS slitmasks that make up the TREX Survey. The position angle (PA) of the long axis of the slitmask is measured counterclockwise from north. Exposure times marked with an asterisk do not represent the effective exposure time and include exposures with bad seeing. For primary targets, "range" indicates main sequence (MS), helium burning (HeB), asymptotic giant branch (AGB), and red giant branch (RGB) stars. \label{tab:masks_info}}
\end{deluxetable*}

\par The DEIMOS spectra were reduced with the {\tt spec2d} and {\tt spec1d} programs \citep{cooper2012, newman2013}. This software has been adapted from its original use for the Sloan Digital Sky Survey to be used on DEIMOS spectroscopy. The resulting one-dimensional spectra were flat-fielded and sky subtracted and then cross correlated against stellar template spectra to measure the line-of-sight velocity of the target star \citep{simon2007}. The velocity measurements were confirmed visually using the {\tt zspec} software. At this step, each measurement is given a quality rating, and rare stars and MW foreground stars are identified (more details below). We then shifted the velocities to the heliocentric frame. 

\par We account for possible miscentering of the star in the slit width direction, which causes a systematic wavelength shift. To do so, we calculated the wavelength shift of the atmospheric absorption line at 7600 \AA. We call this shift the A-band correction and applied it to the measured velocity of the star. We found that the A-band correction varies across the mask depending on the slit's position along the length of the mask, possibly because of a slight positional and/or rotational misalignment of the mask in the sky. To account for the spatial variation, we fitted a polynomial to the A-band velocity as a function of mask position for the stars with the best quality spectra. The polynomial was then applied to all stars to calculate the A-band correction based on the stars' positions along the mask. A typical A-band correction is $\sim -1.3$\kms\ and varies by $\sim 7$\kms\ across a mask.

\par The systematic uncertainties for the old stars observed with the 600 line mm$^{-1}$ and 1200 line mm$^{-1}$ gratings were calculated as in \cite{simon2007, Collins2011}, giving 5.6 \kms for the 600 line mm$^{-1}$ and 2.2 \kms for the 1200 line mm$^{-1}$ grating. We also estimate random uncertainties, derived from duplicate velocity cross correlation measurements of RGB stars (1.65 \kms for the 600 line mm$^{-1}$ and 1.85 \kms for the 1200 line mm$^{-1}$ grating). The final error is the result of adding the estimated random uncertainties to the systematic uncertainties in quadrature. We do not yet have enough duplicate observations of young and intermediate age stars to calculate an estimate of the systematic uncertainty. Initial analysis of the duplicate young stars suggest that a typical velocity measurement error for these stars is $\sim 12$ \kms. For now, we take the velocity errors from the {\tt spec1d} pipeline \citep{cooper2012, newman2013} for the young and intermediate age stars. A typical velocity error measurement is assumed to be  $\sim 6$ \kms, but this is likely an underestimate of the true velocity uncertainty. We will rectify this in the future after obtaining a larger sample of repeat observations. 

\par
\begin{figure*}
    \centering
    \includegraphics[width=\textwidth]{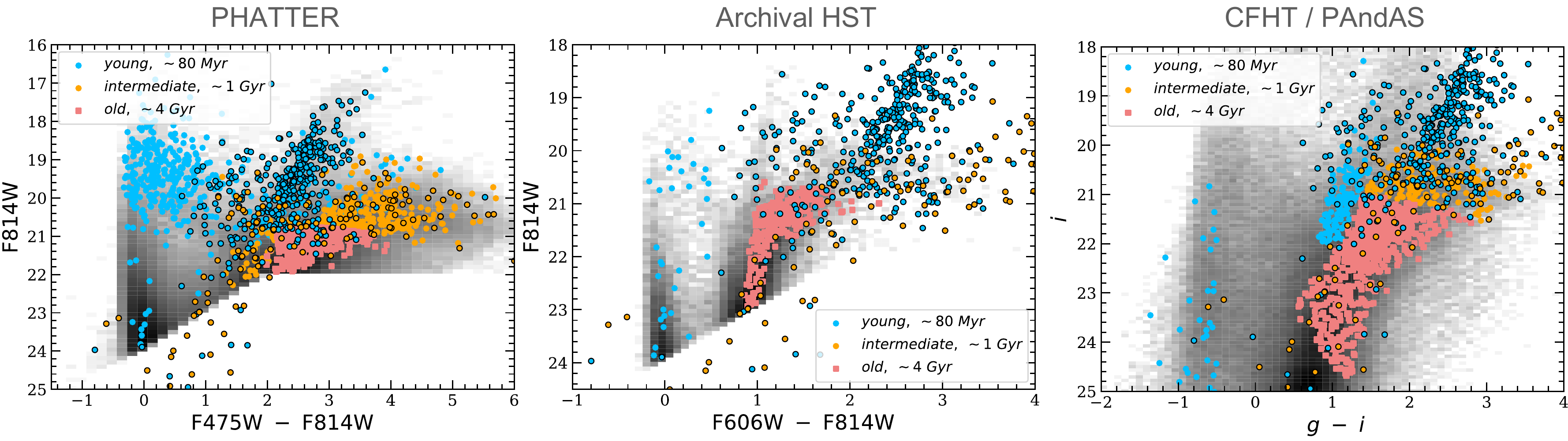}
    \caption{Color-magnitude diagrams of the subsets of photometric catalogs used for target selection with the final stellar sample overplotted. The left panel shows stars with HST photometry with the F457W and F814W bands. Most of these stars are from the PHATTER survey; some are from archival HST images. The center panel shows stars selected from archival HST photometry with the F606W and F814W bands only. The right CMD shows stars selected from the CFHT/Megacam + MegaPipe and from the PAndAS catalogue. In each panel, the blue points represent the young stars, the orange represents the intermediate age stars, and the red squares represent the old stars, and the grey represents a Hess diagram version of the full photometric catalogs. The points outlined in black are young weak CN stars (blue) and intermediate age carbon stars (orange) with ages derived from spectroscopic analysis instead of CMD divisions; see Appendix \ref{sec:rare} for more details. We list the adopted average age for each bin in the legend. All magnitudes have been extinction corrected.}
    \label{fig:cmds}
\end{figure*}

 MW foreground stars are not identified during target selection. Instead, they are removed from our sample if there is Na\,{\smcap i} absorption present during visual inspection of their spectra, as this indicates the star is likely a dwarf star \citep{Gilbert_2006}. Once these visually classified foreground stars are removed, we compared the line-of-sight velocities of the remaining stars to a Besancon model \citep{robin2003,robin2014,robin2017,amores2017} at the location of the TREX Survey with a color and magnitude cut similar to the range of the targets, shown in Figure \ref{fig:cmds}. The MW foreground stars have radial velocities that peak at $-39$ \kms, with $57\%$ of the distribution at $> -50$ \kms. Only 18 stars, or $0.70\%$ of our {\it{final}} sample (described in Section \ref{sec:ages}) has line-of-sight velocities $> -50$ \kms, so our study is largely free of MW contamination.   

\par Targets are also removed from our sample if their spectra suggests the target is an extended source (i.e., background galaxy) or if the quality of the spectrum is too poor to return a well measured velocity. We also eliminate stars with $|v_{LOS}| > 500\rm kms^{-1}$ or that are miscentered in the slit enough to cause a needed correction on the order of $80\rm kms^{-1}$ \citep{sohn2007, simon2007} based on the polynomial fit estimate. Stars with extreme velocities compared to M33's systematic velocity are unlikely to be members of M33 or to have properly measured velocities. With foreground stars, poor quality targets, and duplicate measurements removed, our spectroscopic dataset consists of $4118$ stars in M33. In this specific study, we further narrow our sample based on velocity measurement error (Section \ref{sec:velocities}), CMD placement (Section \ref{sec:ages}), and, for the older stars, the probability of belonging to the stellar disk component (Section \ref{sec:halo}). With the final sample of resolved spectroscopy, we can examine the line-of-sight velocity for individual stars across the disk in M33.

\subsection{Gas Data}

\par Unlike stars, which retain their kinematic memories of past dynamical heating, dense gas's comparatively rapid cooling dissipates energy, leaving the gas as a low velocity dispersion tracer of the disk's gravitational potential. In this study, we use velocity measurements of \hi, CO, and H$\alpha$ to make comparisons between the dynamics of the gas and stars in the disk of M33. 
\par The \hi\ measurements from the Very Large Array (VLA) are described in \cite{gratier2010} and have 5\arcsec--25\arcsec\ resolution (FWHM), depending on the specific pointing (see their Table 2 and 3). The data are from archival VLA imaging obtained in 1997, 1998, and 2001. The resulting gas line-of-sight velocities have an RMS uncertainty of $\sim 1.3$\kms.

\par The CO(2-1) data were observed using the Institute for Radio Astronomy in the Millimeter Range (IRAM) 30 meter antenna by \cite{gratier2010, druard2014}. The observations began in 2008 and build off of those from \cite{gardan2007}. The angular resolution of the data is 12\arcmin\ with a spectral resolution of 2.6 \kms\ (RMS).
\par The H$\alpha$ measurements were observed by \cite{kam2015} at Observatoire du Mont Megantic with a 1.6-m telescope and the Fabry-Perot interferometer in September 2012, producing an angular resolution of $\le 3$\arcsec\ and a typical velocity measurement uncertainty is $\sim 10$\kms\ (FWHM). 

\par Further details on the observations and data reduction are described in the references listed above. All of the gas velocity measurements have been shifted to the heliocentric frame. Each star corresponds to a specific a single pixel of the gas maps, which allows us to make local and direct comparisons of the gas and stellar kinematics. In Section \ref{sec:smoothing}, we discuss how we locally average the stellar kinematics to better match the resolution of the gas imaging so that one star does not represent the stellar kinematics of an entire pixel in a gas map. 
\par The extent of the \hi\ dataset is vast, so we are able to pair almost all stars with the nearest (in projection) \hi\ velocity measurement. The percentages of stars paired with \hi\ measurements are $91\%, 79\%$, and $71\%$ for the young, intermediate age, and old stars, respectively. (We discuss the division of stars into age bins in Section \ref{sec:ages}). The CO and H$\alpha$ velocity measurement maps have a smaller extent so only stars with a projected radius of $\le 4$ kpc are able to be paired to a CO and H$\alpha$ measurement. For H$\alpha$, this includes $24\%, 12\%$, and $4\%$ for the young, intermediate age, and old stars. For the CO, it covers $34\%, 25\%$, and $12\%$ for the young, intermediate age, and old stars.

\section{Broad Age Groups}\label{sec:ages}
We divide stars loosely into three age groups based on average stellar age at present day. First we use color magnitude diagrams (CMD). Different regions on a CMD are dominated by stars of different masses and ages. For example, the MS-dominated region we target consists almost entirely of massive stars with short stellar lifetimes. Regions dominated by evolving AGB stars are populated by intermediate mass stars with present day ages older than the main sequence, but not as old as the targeted RGB stars, which occupy a region dominated by older low mass stars with present day ages $>2.5$ Gyr. 
 
\par We can use stellar population models to estimate average ages for each CMD stellar region \citep{Williams2014, Williams2021}. In the rest of the paper, we will refer to stars in the AGB region as ``intermediate age," and stars in the RGB region as ``old." We combine MS stars, blue HeB, and red HeB, and into a single broad age group that we will refer to as ``young." Even though the RHeB stars are far to the red, they are put into the young group because we targeted high mass ones with short lifetimes. See Figure \ref{fig:cmds} for the approximate location of each stellar lifetime division for our sample. 

\par After the CMD division, we re-categorize weak CN and carbon stars based their spectroscopic information, regardless of their CMD location. Both the intermediate age carbon and young weak CN stars are identified using a combination of visual inspection and machine classification of stellar spectra; they are discussed in greater detail in Appendix \ref{sec:rare}. We assign weak CN stars to the young age group and carbon stars to the intermediate age group because the average age of these stars are consistent with the young and intermediate age group, respectively. We have marked these stars in Figure \ref{fig:cmds} with black outlines to distinguish them from the CMD divisions.

\par We assign each broad bin an average age using simulated CMDs: $\sim 80$ Myr for the young stars group; $\sim 1$ Gyr for intermediate age stars; and $\sim 4$ Gyr for the old stars. These age averages come from \cite{Williams2021, smercina2022} who compare the PHATTER targets to simulated CMDs using Padova isochrones \citep{marigo2017}. These age ranges are quite broad. The range ($16^{\rm th}$ to $84^{\rm th}$) in the young group is $\sim 20-180$ Myr, the intermediate age bin spans ages $\sim 0.56-2.2$ Gyr, and the old age bin spans $1.4-9$ Gyr. (See \cite{Williams_2009} for specific star formation histories of regions in M33.) Additionally, these age bins have some overlap and contamination due to the approximate nature of CMD divisions. However, the average ages for each bin are distinct enough to broadly study stellar kinematics as a function of age, which is the goal of this work. We compare the dynamics of these three broad groups and look for trends with stellar age.

\subsection{Removing Halo Contamination}\label{sec:halo}
\cite{Gilbert2021} provide evidence for the existence of a dynamically hot component in M33, using a subset of the spectroscopic dataset used in this paper. They do not find evidence for this component in their young stellar sample, made up of weak CN stars, which are best described by a single dynamically cold component. The stars of the hot component make up $\sim 22\%$ of the total old sample of the TREX Survey, which we correct for using the model described in \cite{Gilbert2021} to remove likely halo contaminants from the old disk population. 

\begin{figure}[h!]
    \centering
    \includegraphics[scale=.7]{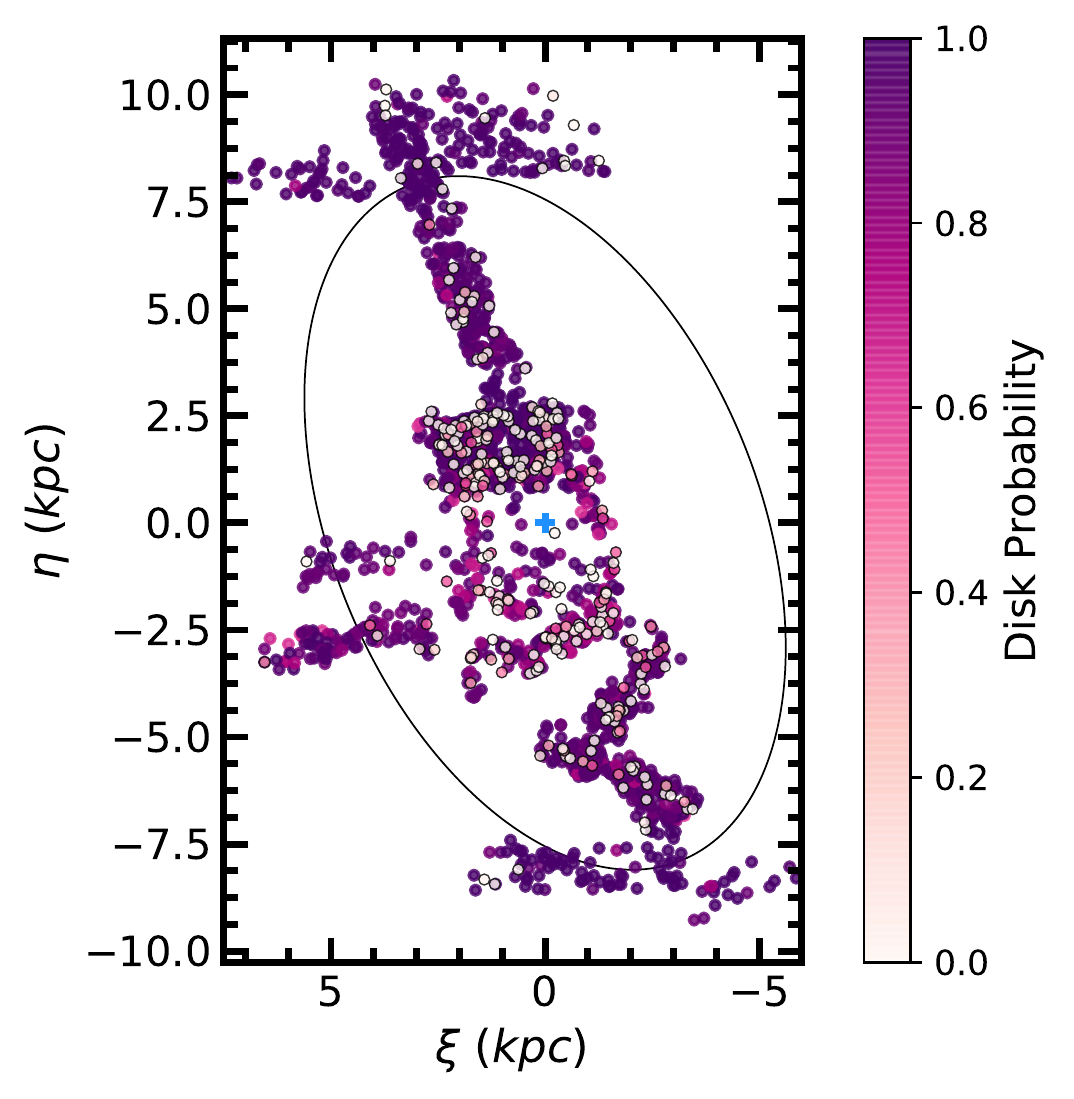}
    \caption{Map of the intermediate age and old stars color coded by probability of belonging to a dynamically cold component. The ellipse represents the approximate location of the disk break. The center of M33 is marked with a blue cross.}
    \label{fig:halo_map}
\end{figure}

\par \cite{Gilbert2021} model the disk and halo assuming a tight kinematic connection to the \hi\ disk. They compare the line-of-sight velocities of stars to the line-of-sight velocity of the \hi\ at the same radius using the titled ring model in \citet{kam2017}, rather than individual \hi\ measurements.
They model the disk and halo as Gaussians in a transformed line-of-sight velocity space defined as the difference between a star's velocity and the calculated line-of-sight velocity of the disk or halo component at the star's disk location, assuming the fractional rotation speed for that component.
This allows each component to rotate at a fraction of the speed of the \hi\ disk model. The best fit model from \cite{Gilbert2021} then returns a probability that a given star's kinematics belong to a dynamically cold component. Although the \citet{Gilbert2021} analysis focuses only on the old stars, we see preliminary evidence that the intermediate age population may host a similar component, and thus we apply the same model to remove candidate kinematically hot AGB stars. The model was run separately on the AGB stars utilizing the same model formalism and procedure used by \citet{Gilbert2021} for RGB stars. We use the probability from the model to keep all intermediate age and old stars with velocities that are at least $80\%$ likely to belong to the dynamically cold component, eliminating velocity outliers and producing a more pure disk-like population. 
We assume all young stars are disk stars. Figure \ref{fig:halo_map} shows a map of the intermediate age and old stars color coded by probability their kinematics are consistent with a cold component.

\par Removing stars with disk probabilities which are $< 80\%$ eliminates $\sim 14\%$ of the initial intermediate age bin and $\sim 23\%$ of the initial old age bin. For the old stars, the percentage of stars eliminated from the disk sample is consistent with the expected fraction of RGB halo stars from \cite{Gilbert2021}. Future work, utilizing an increased AGB sample, will characterize the kinematics of the AGB population as a whole and explore the nature of the AGB stars which have velocities well removed from the M33 disk velocity. In Section \ref{sec:illustris} and \ref{sec:LG}, we explore the implications of not removing possible halo stars.
\par With the quality cuts and the elimination of old halo stars and intermediate age halo star candidates, our study consists of 952 young stars, 521 intermediate age stars, and 1088 old stars for a total of 2561 stars. 

\section{Stellar Kinematics as a Function of Age}
We get line-of-sight velocities for the stars in our sample from the stellar spectroscopic observations. In this section, we describe how we use the line-of-sight velocities to calculate local line-of-sight velocity dispersion, construct rotation curves, and calculate asymmetric drift for each stellar age bin. 

\label{sec:velocities}
\subsection{Local Velocity Smoothing}\label{sec:smoothing}
\begin{figure*}
    \centering
    \includegraphics[scale=.85]{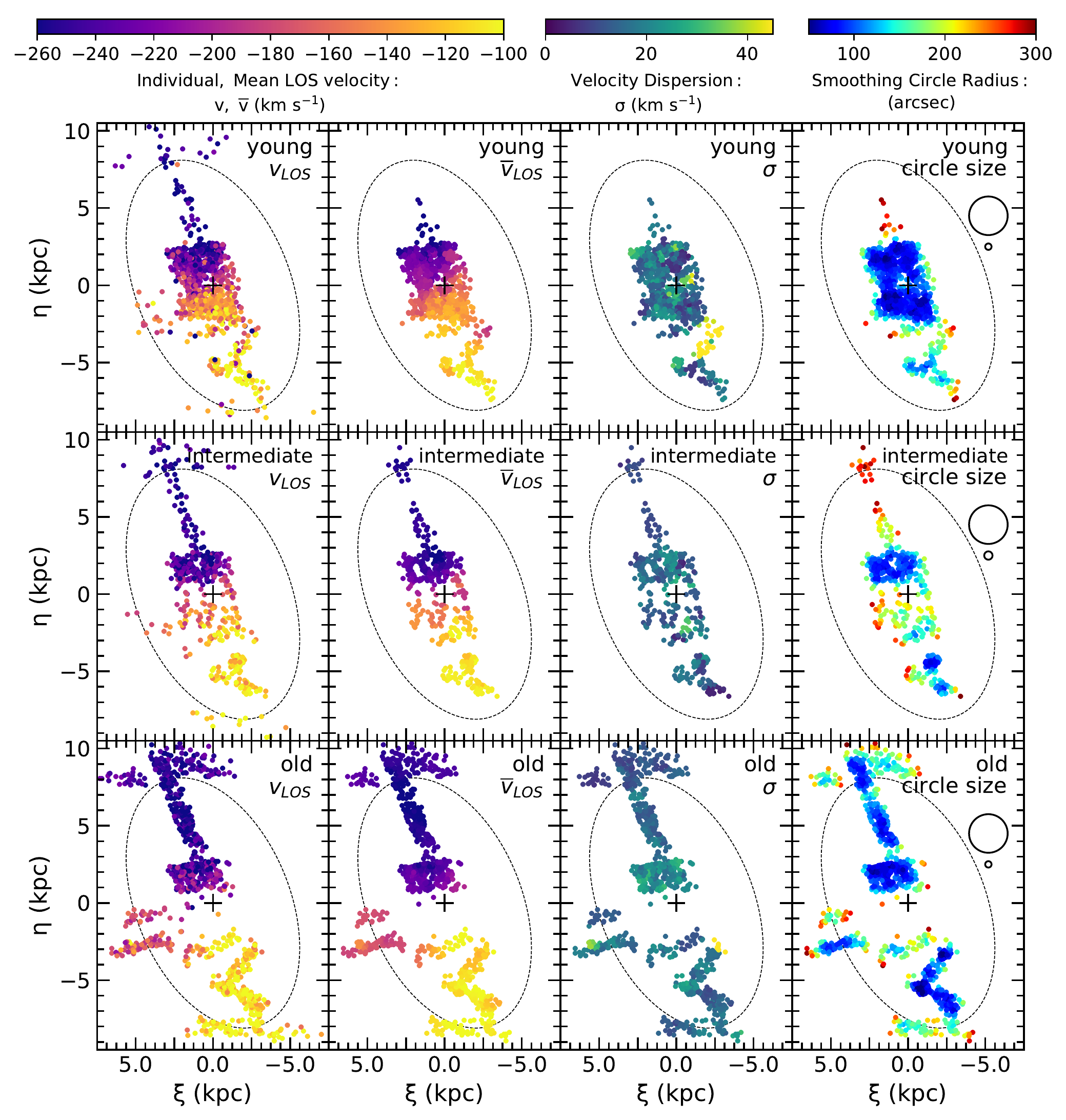}
    \caption{Line-of-sight velocity and velocity dispersion as a function of position for the three broad age bins. The top row shows the young stars. The middle represents the intermediate age population, and the bottom row shows the old population. For all rows, the color in the first column represents individual line-of-sight velocity. For the second column, color represents the smoothed line-of-sight velocity. The third column shows the local velocity dispersion. The last column shows the size of the radius of the smoothing circle used at that point. The smallest and largest smoothing circle used are also shown. The ellipse represents the approximate location of the disk break. The center of M33 is marked with a black cross. The inner disk (r $<3.75$kpc) has higher velocity dispersion for all age bins, and the young stars show an extended area of high dispersion.}
    \label{fig:LOS_map}
\end{figure*}

We calculate the local velocity dispersion by examining the velocities of neighbors around each star. We start with a $50\arcsec$ radius aperture for selecting neighbors and then grow in radius by $5\arcsec$ until there are at least fifteen stars of the same broad age bin within the circle. The sizes of the circles used are illustrated in the last column of Figure \ref{fig:LOS_map}. For the young group, the median radius was $85\arcsec$, for the intermediate age group the median radius was $120\arcsec$, and for the old age group the median radius was $100\arcsec$, which is 0.35 kpc, 0.5 kpc, and 0.42 kpc, respectively, at the distance of M33.

If the circle gets to $300\arcsec$ and still does not contain fifteen members, we do not calculate a line-of-sight velocity average or velocity dispersion at that location. The velocity of the skipped star can still contribute to the analysis of its neighbors however. This smoothing technique is similar to that used in \citet{Dorman2015}, \citet{Quirk2019}, and \citep{Quirk2020}. After smoothing, we have local velocity dispersion measurements centered on 879 young stars, 462 intermediate age stars, and 1053 old stars. 

\par The resulting velocity maps are shown in Figure \ref{fig:LOS_map}. The second column shows the locally averaged line-of-sight velocities for each age bin. The third column shows the local velocity dispersion, the calculation of which we describe below. The fourth column shows the size of the smoothing circle that was used for each center, along with the size of the smallest and largest smoothing circle used for that age bin.

\subsection{Velocity Dispersion}
We calculate the weighted mean of the line-of-sight velocities and the inverse variance weighted root mean square dispersion (RMSE) of the members (Equation \ref{eq:m} and \ref{eq:rmse}). In these two equations, the weights are normalized and derived from the velocity measurement errors ($\sigma_{\rm err, \it i}$ in the following equations) such that, $\sum_{i=1}^{n} w_{i} = 1$ and $w_i = \sigma_{\rm err, \it i}^{-2}$. The velocity measurement error model is discussed in Section \ref{sec:spec}. 

\begin{equation}\label{eq:m}
    \overline{x} = \sum_{i=1}^{n} x_{i} \times w_{i}
\end{equation}

\begin{equation}\label{eq:rmse}
    \sigma_{v} = \sqrt{\sum_{i=1}^{n} (x_{i} - \overline{x})^{2} \times w_{i}}
\end{equation}\label{eq:mean}

We use the inverse variance weighted RMSE as the dispersion. We only consider stars of the same age bin in the averaging and dispersion calculation. 

\par Along with local velocity dispersion, we compute the median velocity dispersion for each of the three age bins, which is reported in Table \ref{tab:LOS_disp}. We can compare these values to the global models fit from \citet{Gilbert2021}, who find a global velocity dispersion of $\sim 16$\kms\ for the young stars and $\sim 21$\kms\ for the old stars, which is similar to the median local velocity dispersions reported here. 

\par
\begin{figure}[bp!]
    \centering
    \includegraphics[width=\columnwidth]{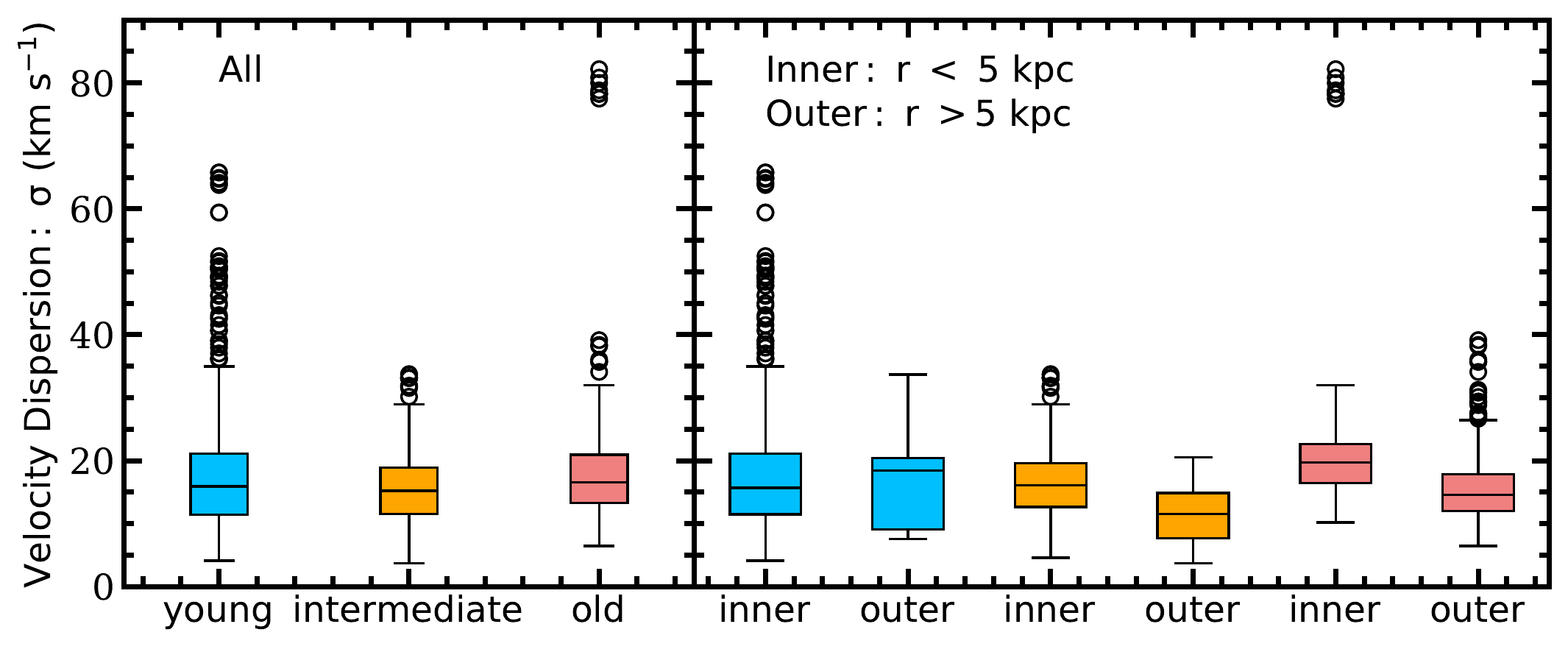}
    \caption{Velocity dispersion distributions for the three age bins. The left panel shows the distributions for the full final population. The right panel shows each age bin divided into an inner ($r\ <\ 5$ kpc) subgroup and an outer subgroup ($r\ >\ 5$ kpc). For all boxplots, the shaded box represents the interquartile range, the horizontal line across the box represents the median, and the open circles show the outliers. The outliers are stars with a velocity dispersion that are at least 1.5 $\sigma\ $ from the first or third quartile value, which is the distance marked by the whiskers. Velocity dispersion does not vary significantly with stellar age and is on average higher in the inner region.}
    \label{fig:disp_box}
\end{figure}

 We also show details of the velocity dispersion distributions in Figure \ref{fig:disp_box}. The left panel shows the median value, interquartile range, and outliers for the three age bins across the full extent of the TREX Survey. Overall, we find that velocity dispersion does not vary strongly with stellar age, as the median values of each population do not vary significantly. Furthermore, the median values of each age bin are relatively low and are roughly twice the average dispersion of the \hi\ \citep[$8$ \kms]{chemin2020}. The low magnitude of velocity dispersion and the lack of an increase with stellar age are not expected when compared to expectations from simulations of slightly more massive disk galaxies \citep{martig2014} or observations of star clusters in M33 \citep{beasley2015}. However, these studies do not remove dynamically heated populations that are likely to belong to a halo, so it is not an exact comparison.

\par \citet{martig2014} find that the shape of the age-velocity dispersion relation for young and intermediate age stars is dependent on recent merger or other heating activity, whereas the velocity dispersion for the oldest stars is more dependent on birth kinematics. They also find that uncertainties in ages can obscure possible trends in the age-velocity dispersion relation. Since the age bins in this work are broad, we could be missing a more subtle trend, but this would not explain the low magnitudes.

\par While the medians of the distributions are similar, the distributions themselves are broad enough that they may be widened by trends with radius. In the right panel of Figure \ref{fig:disp_box}, we show the distributions of velocity dispersion for each age group broken into an inner ($r\ <\ 5$ kpc) and outer ($r\ <\ 5$ kpc) subgroup. For all age bins, the distribution of velocity dispersion shifts lower in the outer region compared to in the inner region, although the median velocity dispersion for the young stars is higher in the outer region. The number of outliers is higher in the inner region for the young and intermediate age stars. 

\begin{table}[]
    \centering
    \begin{tabular}{c|c}
    \hline
    Age Group & Mean $\sigma_{LOS}\ \rm (kms^{-1})$ \\
    \hline
         Young & $15.9_{-0.2}^{+0.3}$ \\
         Intermediate Age & $15.2_{-0.2}^{+0.2}$ \\
         Old & $16.5_{-0.1}^{+0.2}$ \\
    \hline
    \end{tabular}
    \caption{Medians of weighted velocity dispersion as a function of broad age bin. The errors on the median value represent the difference between the $16^{\rm th} \rm and\ 84^{\rm th}$ percentiles divided by $\sqrt{\rm N}$, where N is the number of stars.}
    \label{tab:LOS_disp}
\end{table}

\par
\begin{figure*}
    \centering
    \includegraphics[width=\textwidth]{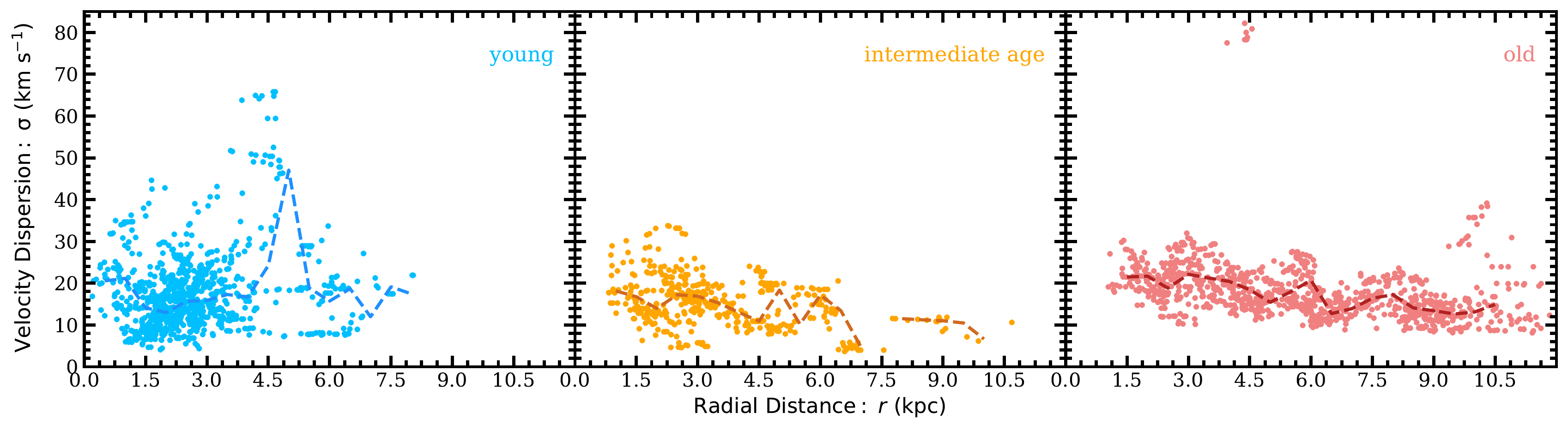}
    \caption{Velocity dispersion as a function of deprojected radius for the young, intermediate age, and old populations. Median lines are also plotted. For the intermediate and old age bins, velocity dispersion is higher in the inner regions than in the outer regions. The old stars show one concentrated area of extreme velocity dispersion, while the young stars show an extended area of high dispersion, and the intermediate age stars show no high velocity dispersion. The area of extreme velocity dispersion for the old stars is overlapped with an area that also has extreme velocities for the young stars.}
    \label{fig:disp_dist}
\end{figure*}

 We look more closely at velocity dispersion as a function of radius in Figure \ref{fig:disp_dist}, which shows the distributions of velocity dispersion and radius. There is a clear downward trend in the intermediate age and old stars populations as one goes out to greater radii. This suggests the outer disk is dynamically cooler than the inner disk, which is consistent with the findings of \citet{Gilbert2021} and disk galaxies in general \citep[e.g.][]{bottema1993}. \citet{koch2018} find that the velocity dispersion of the \hi\ decreases by $\sim 2$ \kms from the center of M33 to $\sim r = 8$ kpc. For the young stars, any trend is less clear because of the extended area of high dispersion, which we examine below, however there is a slight trend of increasing velocity dispersion with radius. 

\begin{figure*}
    \centering
    \includegraphics[width=\textwidth]{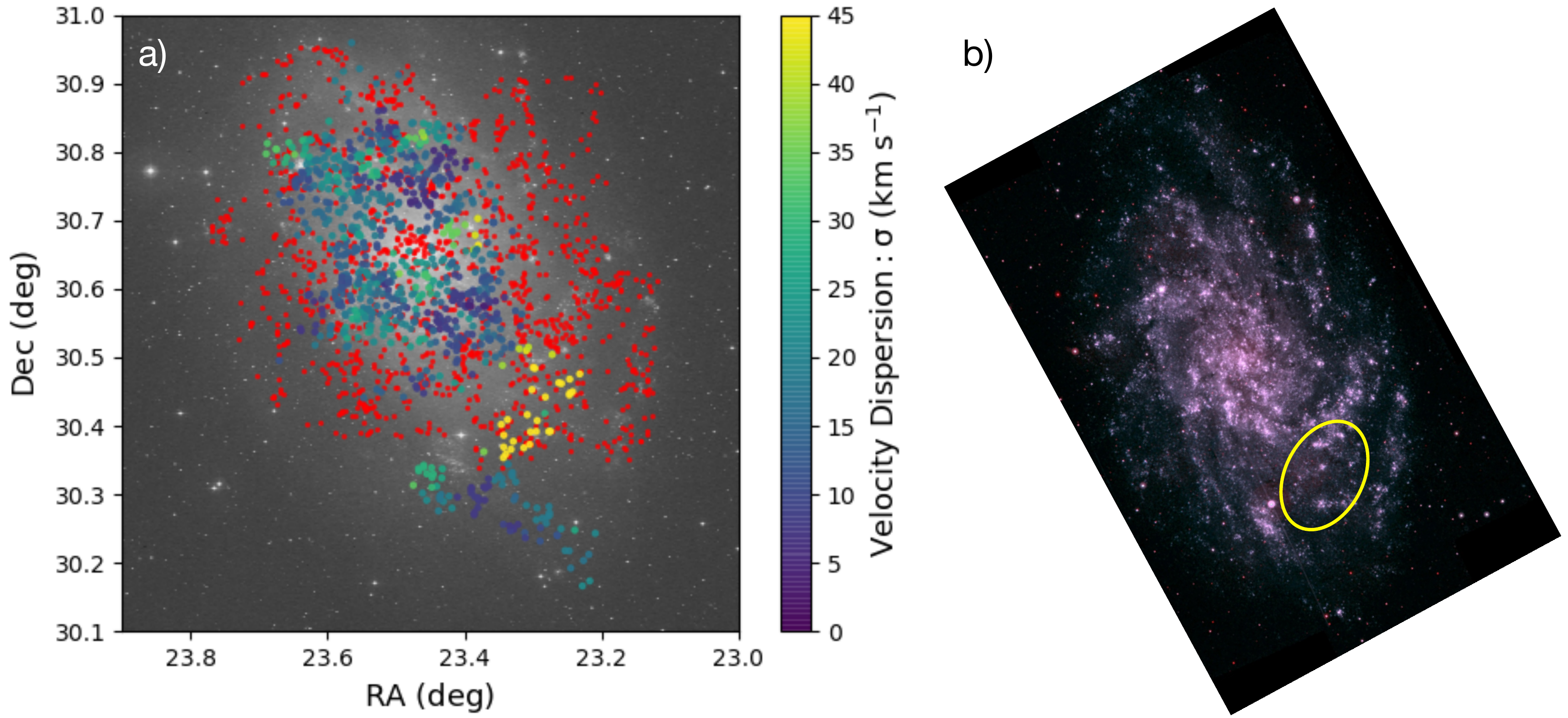}
    \caption{Velocity dispersion as a function of position for the young stars. Panel a) shows the young stars overplotted on an image of M33 from the DSS. The color of the points corresponds to velocity dispersion. The smaller red points mark H\,{\smcap ii} regions \citep{hodge1999}. Panel b) shows a UV image of M33 from the Swift Observatory (image credit: NASA/Swift/Stefan Immle) with a yellow ellipse showing the approximate location of the high velocity dispersion. (The DSS map is slightly enlarged compared to the image of M33 on the right.) These locations do not have higher concentrations of UV emission than other parts of the disk.}
    \label{fig:young_disp}
\end{figure*}

\par Although there is not much trend between velocity dispersion and age, there is an extended area of extreme velocity dispersion for the youngest group of stars. 

\noindent Figure \ref{fig:LOS_map} shows this extended area of high velocity dispersion in the young star population. We examined this anomalous region in detail to test if the substructure is real or potential velocity measurement artifact. First, we looked at young stars with extreme velocities that contribute to the high velocity dispersion in this extended region. Their velocities are well measured and pass our quality cuts.
Secondly, this region is also not dominated by the largest size smoothing circle. Because of this, we think the areas of high velocity dispersion in the young stars are real effects and will investigate it further using duplicate velocity measurements as future work. A smaller portion of this same area is also coincident with high velocity dispersion for the old stars that do have well characterized velocity measurement errors. 

\par We also compare the location of the high velocity dispersion region to a UV map of M33 in Figure \ref{fig:young_disp}. The yellow ellipse in the right panel shows the approximate location of the extended area of high velocity dispersion with respect to the UV emission. Because there is ongoing star formation across the disk, there does not appear to be anything notable about the specific high dispersion area and no correlation between stellar velocity dispersion and whether a location is UV dark or UV light. The left panel shows velocity dispersion as a function of position compared to H\,{\smcap ii} regions from \citet{hodge1999}. There also does not seem to be anything coincident about the H\,{\smcap ii} regions around the high stellar velocity dispersion.

\par \citet{koch2018} examine the atomic ISM across the disk of M33 and find a filament of \hi\ that extends to 8 kpc across the southern disk (see their Figure 18). While the filament overlaps the area of high velocity dispersion reported here, the young velocity dispersion is coincident with a void in the filament. It is further unlikely that the filament is related to the high stellar velocity dispersion because the line-of-sight velocities of the young stars would need to be blueshfited by $>30$ \kms, which they are not. There are some localized bright areas of \hi\ with broad line widths that are close to the area of high velocity dispersion shared by the young and old stars, but it is unclear if they are related to the high stellar dispersion.

\par It is possible that other phenomenon are causing the high velocity dispersion in the young stars and small number of old stars. For example, there could be unmarked massive gas clouds or a significant number of stellar binaries. Additionally this area could contain substructure from a relatively recent minor merger that lies in front of the disk. If M31 and M33 did have an interaction in the past, perhaps the interaction could have increased the dispersion of the \hi\ and young stellar disk.

\subsection{Rotation Curves}
We use the weighted average line-of-sight velocities to calculate the rotation velocities of the stars, which we compare to the rotation velocities of gas tracers such as \hi, CO, and H$\alpha$. To calculate the rotation velocity, we convert the line-of-sight velocity to a circular velocity using the tilted ring model described in \cite{kam2017}. \cite{kam2017} divide the \hi\ disk of M33 into forty-nine annuli from $r=0\arcmin$ to $r=96\arcmin$ and measure the position angle (PA) and inclination ($i$) of each annulus, as tabulated in Table 4 of their paper. The rings in \citet{kam2017} have a width of $0.2'$. We interpolate this table to create thinner rings. We then calculate a star's deprojected distance from M33's center using the PA and $i$ of M33. With that distance, we match the star/gas measurement to a tilted ring from the interpolated table and assign it the corresponding ring's PA and $i$. We recalculate the deprojected distance, and reassign the star/gas measurement to another ring if needed. We repeat this process twice before adopting the final PA and $i$ for the star/gas measurement.

\begin{figure}[h!]
    \centering
    \includegraphics[scale=.8]{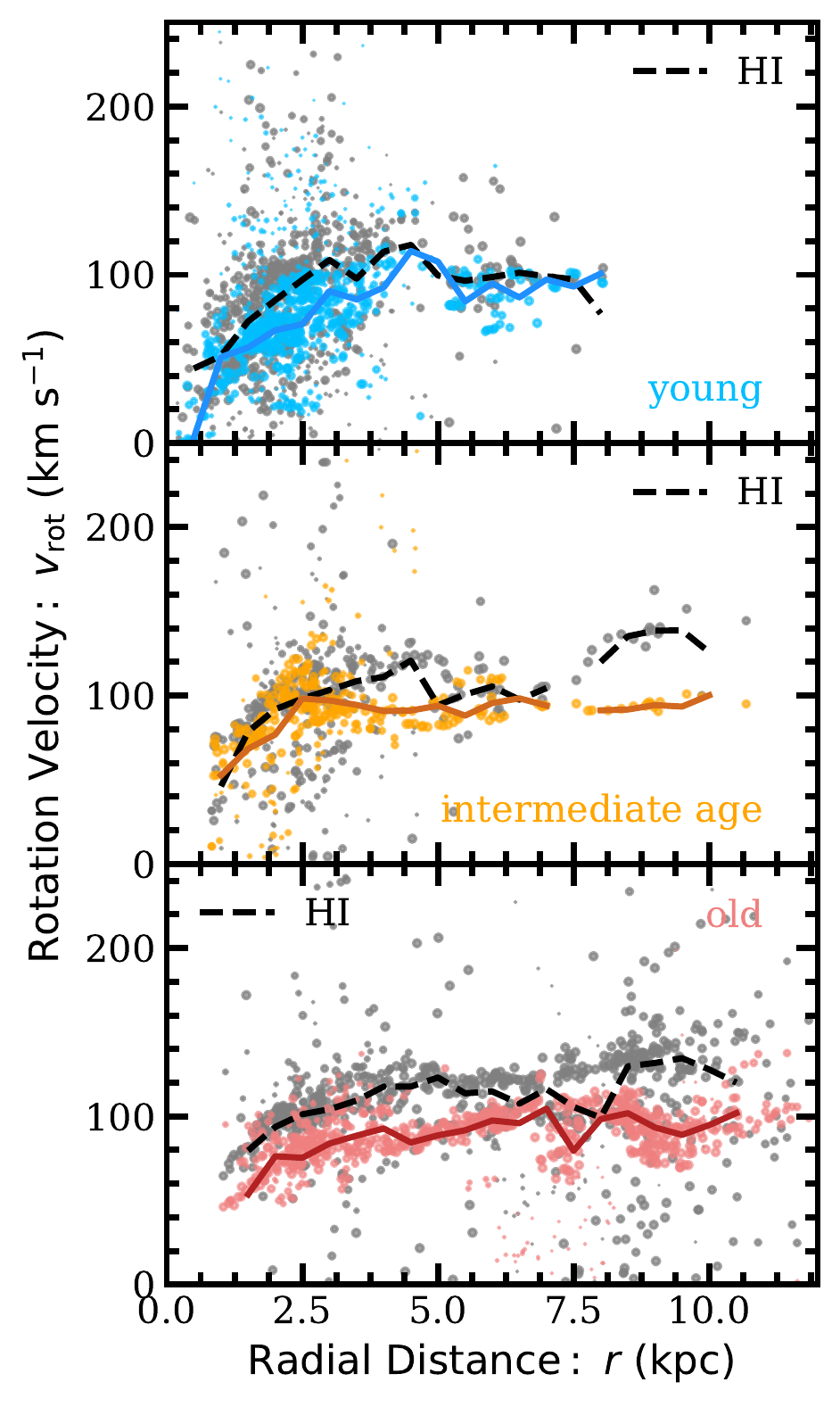}
    \caption{Rotation velocity as a function of deprojected radius. Rotation velocities are calculated with the tilted ring model in \cite{kam2017} (Equation \ref{eq:vrot}). The top panel shows the youngest age bin (light blue), the middle panel shows the intermediate age group (orange), and the bottom panel shows the old group (red). Each star has been paired with a \hi\ velocity measurement along the same line-of-sight, and the gas is represented by the grey dots. The size of the points is proportional to the $\cos(\theta)$ factor in Equation \ref{eq:vrot} to illustrate the limitations of the equation around the minor axis. The solid (dotted) line shows the median rotation velocity for 0.5 kpc bins for the stars (\hi). The deprojection effects around the minor axis cannot explain the full amount of scatter, especially for the gas.}
    \label{fig:rc_full}
\end{figure}

\begin{equation}\label{eq:vrot}
    v_{\rm rot} = \frac{v_{\rm LOS} - v_{\rm sys}}{\cos(\theta)\sin(i_{TR, \star})}
\end{equation}

\begin{figure*}
    \centering
    \includegraphics[scale=.74]{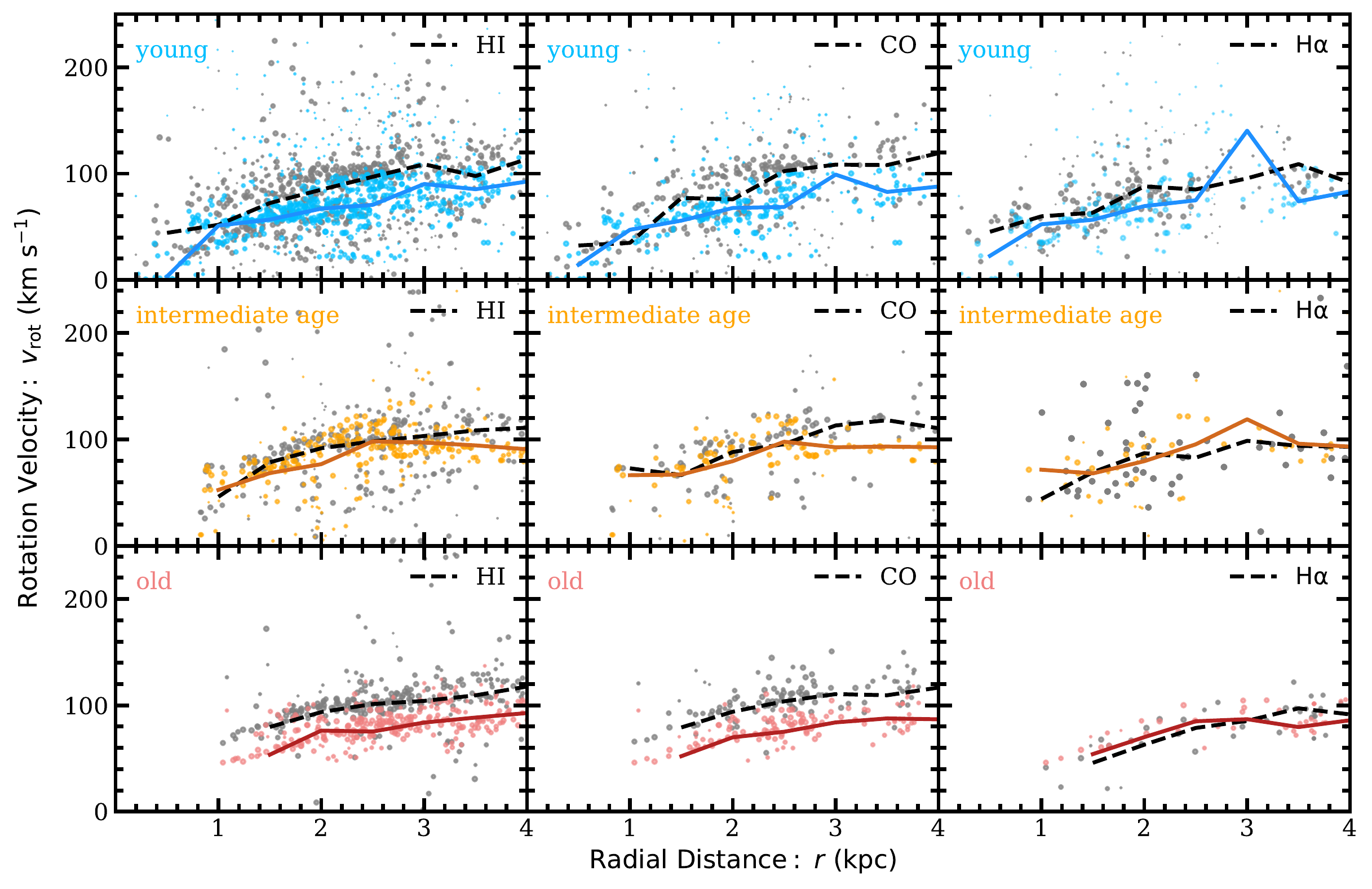}
    \caption{Rotation velocity as a function of deprojected radius for the inner 4 kpc of the disk. Rotation velocities are calculated with the tilted ring model in \cite{kam2017} (Equation  \ref{eq:vrot}). The top row shows the youngest age bin (light blue), the middle row shows the intermediate age group (orange), and the bottom row shows the old group (red). Each star has been paired with a \hi\ (first column), CO (second column), and H$\alpha$ (third column) velocity measurement along the same line-of-sight, and the gas is represented by the grey dots. The size of the points is proportional to the $\cos(\theta)$ factor in Equation \ref{eq:vrot} to illustrate the limitations of the equation around the minor axis. The solid (dotted) line shows the median rotation velocity for 0.5 kpc bins for the stars (gas). The deprojection effects around the minor axis cannot explain the full amount of scatter, especially for the gas. The bar is believed to extend to 0.5 kpc \citep{Williams2021, smercina2022, lazzarini2022}.}
    \label{fig:rc_inner}
\end{figure*}

The above equation projects a star onto a circular orbit. We use $v_{sys}= -180 \rm kms^{-1}$ \citep{kam2017}. Theta is the azimuthal angle in the plane of M33, and $i$ is the inclination that comes directly from matching a star/gas measurement to a tilted ring. Theta is calculated with $ \theta = \beta \times [\alpha \cos(i_{TR})]^{-1}$ where $\alpha = \eta \cos(\rm PA_{\it TR}) + \xi \sin(\rm PA_{\it TR})$ and  $\beta = \xi \cos(\rm PA_{\it TR}) - \eta \sin(\rm PA_{\it TR})$. We use the assigned PA of the ring that a star/gas measurement lies in and the deprojected coordinates of the star/gas measurement centered on M33. Since each star is paired with a gas measurement, the star and the gas measurement share the same deprojected geometric parameters but can have different values for rotation velocities, as they have different line-of-sight velocities. 

\begin{figure}[tp!]
    \centering
    \includegraphics[width=\columnwidth]{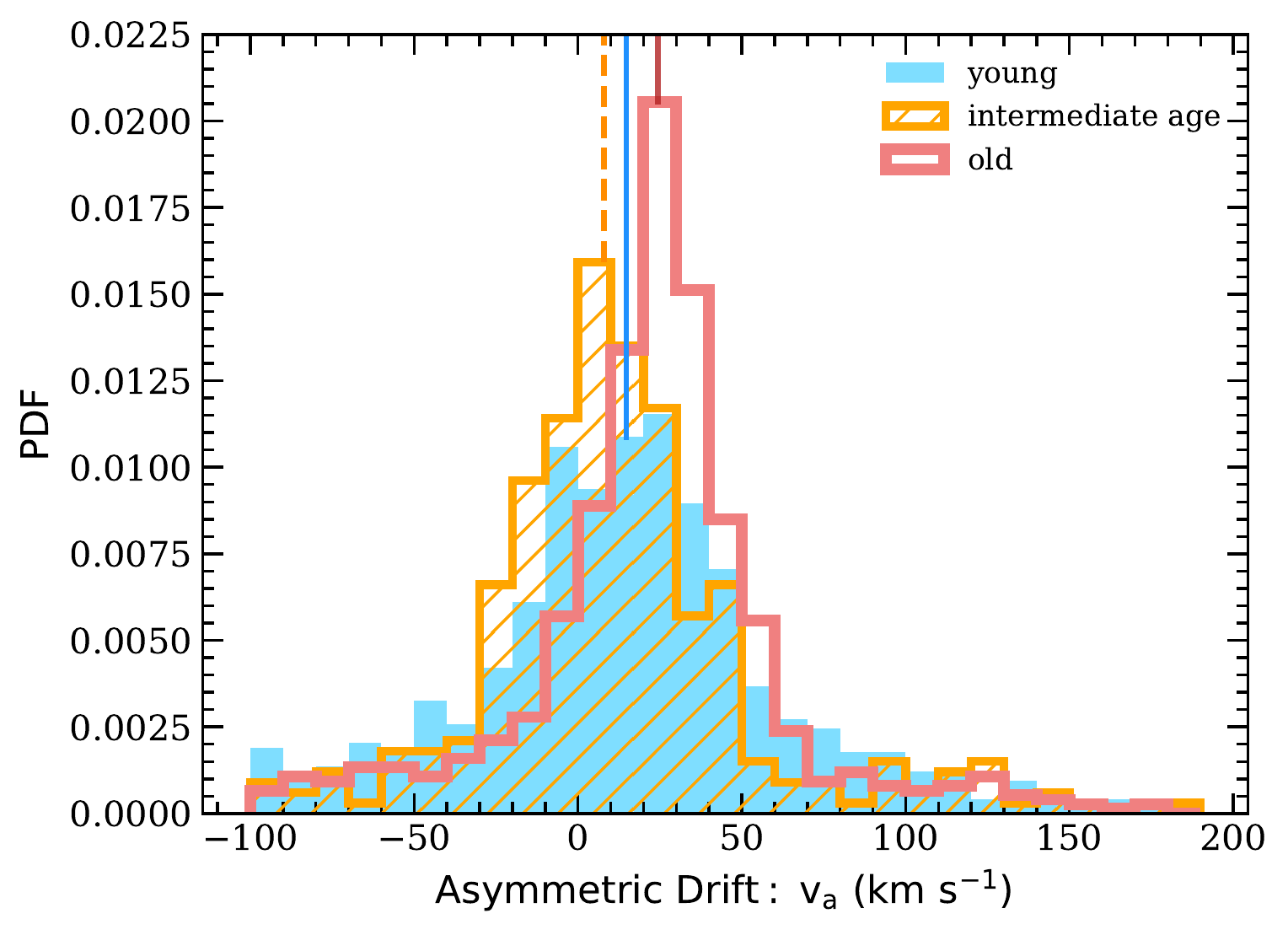}
    \caption{Asymmetric drift distributions for the young stars (blue solid), intermediate age (orange hatched), and old star (red outline). AD is calculated with respect to the \hi\ using $v_a = v_{\rm rot,\ gas} - v_{\rm rot,\star}$. The medians are marked by the vertical lines that run from the peak of the corresponding distribution to the top of the plot. There is no clear trend between AD and stellar age, except that the width of the distribution decreases with stellar age.}
    \label{fig:ad_full}
\end{figure}

With the above equations, we construct a rotation curve for each of our broad stellar age bins. The three rotation curves are shown in Figure \ref{fig:rc_full}. These rotation curves show the rotation velocity as a function of deprojected radius for the three stellar age bins and for the \hi\ along the same line-of-sight of each star for the full extent of the TREX Survey. We also plot the inner rotation curves compared to the CO and H$\alpha$ datasets, which do not extend beyond 4 kpc so are not shown in the full rotation curve. These inner rotation curves are plotted in Figure \ref{fig:rc_inner} for the inner 4 kpc for the three age bins and \hi, CO, and H$\alpha$. 

\par In both Figures \ref{fig:rc_full} and \ref{fig:rc_inner} it is clear that the rotation curves of both the stars and the gas show significant scatter. \cite{Quirk2019} demonstrate the deprojection factor of the tilted ring model, or the denominator, approaches zero along the minor axis, regardless of inclination because of the $cos(\theta)$ factor, which can explain some but not all of the scatter in the rotation curves. Other sources of scatter in the stellar rotation curves could be from poorly measured velocities, particularly in the young stars, or disturbances from M33's bar \citep{Williams2021, smercina2022, lazzarini2022} and spiral arms. Scatter in the gas rotation curves could reflect the impact of star formation or turbulence in the ISM. The amount of scatter in the stellar rotation curves is largest for the youngest group and decreases with stellar age, which suggests M33's high star formation rate could be causing turbulence not only in the gas but also in the birth kinematics of stars born from that gas. This could also be causing some of the extreme velocities in the young stars.

\par To quantify the scatter, we have added median lines to each stellar and gas rotation curve and have made the marker size proportional to the $cos(\theta)$ factor. The scatter cannot entirely be attributed to deprojection effects along the minor axis, especially for the gas, which appears to have the most scatter away from the minor axis. We used binning of 0.5 kpc to remove likely outliers while preserving local discrepancies for the median lines. Any velocity difference between the stellar and gas rotation curves is a visual representation of asymmetric drift and will be explored further in the next section.

\subsection{Asymmetric Drift}

We use the stellar and gas rotation curves in Figures \ref{fig:rc_full} and \ref{fig:rc_inner} to calculate the asymmetric drift (AD or $v_{a}$) of the three broad age bins. Often, AD is defined as the difference between the circular velocity derived from the potential of a galaxy and the rotation velocity of the stars \citep{Stromberg}. For the purpose of this study, we define AD to be the difference between the rotation velocity of the gas and that of the stars \citep{Quirk2019, Quirk2020}. This choice allows us to make local and empirical AD measurements without relying on models of the potential of M33. 

\par AD measurements can be used to measure dynamical heating \citep{Quirk2019}, as gas, which is collisional, can fairly easily dissipate energy and maintain a low energy orbit \citep{Sellwood11999}, while stars retain a non-circular orbit if they have been perturbed onto an eccentric orbit \citep{Leaman2017, Sellwood2002}. In M31, observed AD compared to that in simulated IllustrisTNG-100 M31-like analogs has provided evidence that M31 experienced a relatively recent major merger \citep{Quirk2020}. In our own galaxy, AD is routinely used to correct the local standard of rest and to predict rotation curves outside the solar neighborhood before Gaia \citep{Golubov2014, Huang2015}.

\par
\begin{figure*}
    \centering
    \includegraphics[width=\textwidth]{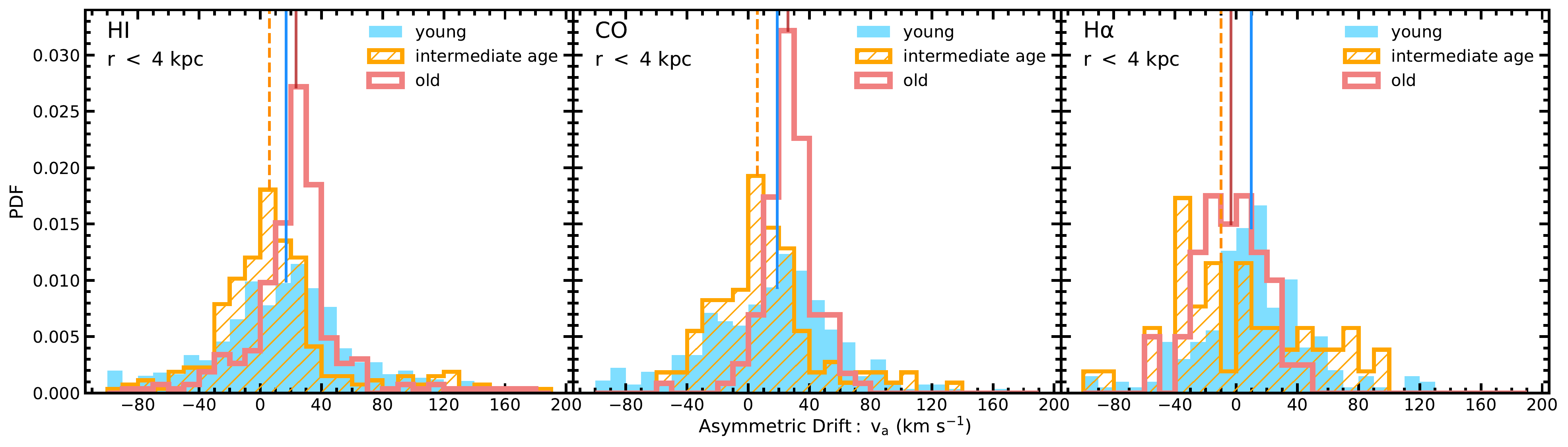}
    \caption{AD distributions for the young stars (blue solid), intermediate age (orange hatched), and old groups (red outline) for the inner 4 kpc. AD is calculated with respect to the \hi\ (left panel), CO (middle panel), and H$\alpha$ (right panel) using $v_a = v_{\rm rot,\ gas} - v_{\rm rot,\star}$. The medians are marked by the vertical lines that run from the peak of the corresponding distribution to the top of the plot. On average, the young and old stars have similar median AD values, suggesting there is no trend between AD and stellar age. For the H$\alpha$, the gas is lagging the intermediate age and old stars, resulting in a negative AD, but both AD values are consistent with zero within 1$\sigma$.}
    \label{fig:ad_inner}
\end{figure*}

 For every star and gas measurement pair, we calculate an AD value, $v_{a}$, from their respective rotation velocities, using $v_a = v_{\rm rot,\ gas} - v_{\rm rot,\star}$. We make AD measurements with respect to \hi, CO, and H$\alpha$. (This equation is in rotation velocity space, whereas the offset used in \cite{Gilbert2021} for halo star identification is in line-of-sight space.) The CO and H$\alpha$ measurements only exist for the inner $\sim4$ kpc so we calculated AD in the inner regions with respect to all three kinds of gas but only with respect to \hi\ for the full survey extent (or out to $\sim 11$ kpc). The distribution of these values are plotted in Figure \ref{fig:ad_full} and \ref{fig:ad_inner}, and the median values, width of the distributions, and percentage of outliers are listed in Table \ref{tab:ad_HI} and \ref{tab:ad_inner}. The outliers are stars with a velocity dispersion that are at least 1.5 $\sigma\ $ from the first or third quartile value.

\begin{table}
    \centering
    \begin{tabular}{c|c|c|c}
    \hline
    Age Group & AD w.r.t \hi\ & Width & Outliers \\
    & ($\rm kms^{-1}$) & ($\rm kms^{-1}$) & \% \\
    \hline 
        Young & $15.1_{-1.7}^{+1.6}$ & 73.3 & 3.2\\
        Intermediate Age & $8.0_{-1.6}^{+1.7}$ & 68.2 & 5.8 \\
        Old & $24.5_{-1.1}^{+0.9}$ & 51.1 & 3.7\\
    \hline
    \end{tabular}
    \caption{Stats for the distribution of AD for the three age bins with respect to \hi\ for the full extent of the survey. Median values of AD, width (sigma) of the distribution, and the percentage of outliers in the distribution are shown. The errors on the median value represent the difference between the $16^{\rm th} \rm and\ 84^{\rm th}$ percentiles divided by $\sqrt{\rm N}$, where N is the number of stars. The outliers are stars with a velocity dispersion that are at least 1.5 $\sigma\ $ from the first or third quartile value.}
    \label{tab:ad_HI}
\end{table}

\begin{table*}[]
    \centering
    \begin{tabular}{c|c|c|c|c|c|c|c|c|c}
    \hline
    Age Group & AD w.r.t \hi\ & Width & Outliers & AD w.r.t CO & Width & Outliers & AD w.r.t H$\alpha$ & Width & Outliers\\
    & ($\rm kms^{-1}$) & ($\rm kms^{-1}$) & \% & ($\rm kms^{-1}$) & ($\rm kms^{-1}$) & \%  & ($\rm kms^{-1}$) & ($\rm kms^{-1}$) & \% \\
    \hline 
        Young & $16.8_{-1.9}^{+1.8}$ & 72.0 & 2.7 & $18.9_{-3.0}^{+2.2}$ & 70.6 & 3.2 & $9.8_{-2.6}^{+2.4}$ & 64.0 & 3.9 \\
        Intermediate Age & $5.9_{-1.7}^{+1.6}$ & 65.7 & 8.3 & $6.0_{-2.7}^{+2.4}$ & 49.0 & 3.6 & $-10.0_{-4.3}^{+8.5}$ & 72.7 & 3.6\\
        Old & $23.3_{-1.4}^{+1.0}$ & 38.1 & 6.4 & $26.0_{-1.3}^{+1.3}$ & 17.4 & 0.9 & $-3.4_{-3.3}^{+3.5}$ & 21.8 & 0.0\\
    \hline
    \end{tabular}
    \caption{Stats for the distribution of AD for the three age bins with respect to \hi, H$\alpha$, and CO for the inner 4 kpc. Median values of AD, width (sigma) of the distribution, and the percentage of outliers in the distribution are shown. The errors on the median value represent the difference between the $16^{\rm th} \rm and\ 84^{\rm th}$ percentiles divided by $\sqrt{\rm N}$, where N is the number of stars. The outliers are stars with a velocity dispersion that are at least 1.5 $\sigma\ $ from the first or third quartile value.}
    \label{tab:ad_inner}
\end{table*}

For the \hi\ and CO, the intermediate stars show the lowest AD values, so there is no monotonic trend with stellar age. When compared to H$\alpha$, the intermediate age stars have the greatest offset in the rotation velocity of the gas and stars; however, this difference is barely larger than 1$\sigma$ so is not robust. Furthermore, the median AD values are negative for the intermediate age and old stars, which means that on average, the H$\alpha$ is lagging behind the stars and therefore the stars could be more settled on a circular orbit than the H$\alpha$. However, both of these AD values are consistent with 0 within 1$\sigma$. We expect AD to increase with stellar age \citep[e.g.,][]{Sellwood2002, Westfall2007, Gaia_Co2021, Quirk2019}, so seeing a high magnitude of AD in the young stars suggests there is something perturbing the young stars on short timescales, which resulted in their kinematics diverging from the gas. 

\par Overall, the old stars tend to have the greatest magnitude of AD, but the distributions of AD values are narrower. In the inner 4 kpc, the young and intermediate age stars have the largest widths, that are at times double that of the old stars for each of the three types of gas used in calculating AD, which is the opposite of what is expected for a canonical model of steady dynamical heating of stars. The old stars are thus more systematically offset from the nearby gas in terms of the kinematics, whereas the young and intermediate stars are more randomly offset from the kinematics of the nearby gas. Over the full radial range, even though the young stars still have the widest AD distribution, the three groups of stars have more similar widths. This suggests the stars in our sample are most similar between radii 4 an 11 kpc.

\par The percentage of outliers (stars with AD values that are $1.5\sigma$ or greater away from the $16^{\rm th}$ or $84^{\rm th}$ percentiles) in each distribution shows no clear pattern with median AD, width of the distribution, or stellar age. The intermediate age stars have the highest percentage of outliers in the distribution of AD values calculated with respect to \hi\ in the inner 4 kpc.

\par It is important to note that the three gas datasets used come from different telescopes and were reduced using different methods. Subtle differences in the derivation of the gas velocity measurement could be causing variations in the AD median values, not just physical phenomena. However, subtle differences would systematically affect the three age bins equally, which means these potential differences cannot be obscuring a lack of trend between age and asymmetric drift.

\par In summary, M33 does not show the expected increase in AD with stellar age because of the low AD of the intermediate age stars and the high AD value for the young stars. The distributions of AD get narrower with stellar age, showing that the younger stars are more randomly offset from the gas than the older stars.

\section{Comparison to IllustrisTNG}
\label{sec:illustris}

\par We analyze M33-like analogs from the IllustrisTNG50-1 cosmological simulation to comment on the uniqueness of the results presented above. The IllustrisTNG Project is a suite of N-body and hydrodynamic simulations from redshift $z=127$ to $z=0$. It uses the \texttt{AREPO} moving-mesh code \citep{marinacci18, naiman18, springel18, pillepich18, nelson18, springel10}. The simulations have a cosmological volume of (51.7 Mpc)$^3$ and have the following initial cosmological parameters from \cite{planck15}: $\Omega_m= 0.3089$, $\Omega_{\Lambda}=0.6911$, $\Omega_{b}=0.0486$,  $\sigma_8=0.8159$, $n_s=0.9667$, and $h=0.6774$. For our analysis, we adopt a value of $h=0.704$ \citep[WMAP-9;][]{hinshaw13}.

\par We use data from the IllustrisTNG50-1 simulation (hereafter IllustrisTNG), the smallest but highest resolution simulation in the project that includes both dark matter and baryons. IllustrisTNG follows the evolution of 2160$^3$ dark matter particles and 2160$^3$ hydrodynamical cells, resulting in a a baryonic mass resolution of $m_{\rm bary} = $ $8.1 \times 10^4 \, M_{\sun}$ and a dark matter particle mass resolution of $m_{\rm DM}=4.4 \times 10^5 \, M_{\sun}$. Halos and subhalos are identified using  \texttt{SUBFIND} \citep{springel01, dolag09}. 

\par We identify general M33-like analogs as halos that fit the following criteria: the subhalo is the central/primary subhalo at $z=0$ in Friend of Friend (FoF) group with a virial mass of $M_{\rm vir}=1-2.5 \times 10^{11}M_{\sun}$ \citep[and references within]{patel18}; the subhalo has a stellar mass of $M_{\star}=2.8-5.6 \times 10^{9} M_{\sun}$ \citep{guo2010}, and the subhalo's maximum dark matter particle circular velocity is $< 70$\kms. These cuts produce a sample of 224 analog galaxies. We eliminated eight of these due to lack of baryon particles across the inner 10 kpc, leaving 216 M33-like galaxy analogs.

\begin{figure*}[]
    \centering
    \includegraphics[width=\textwidth]{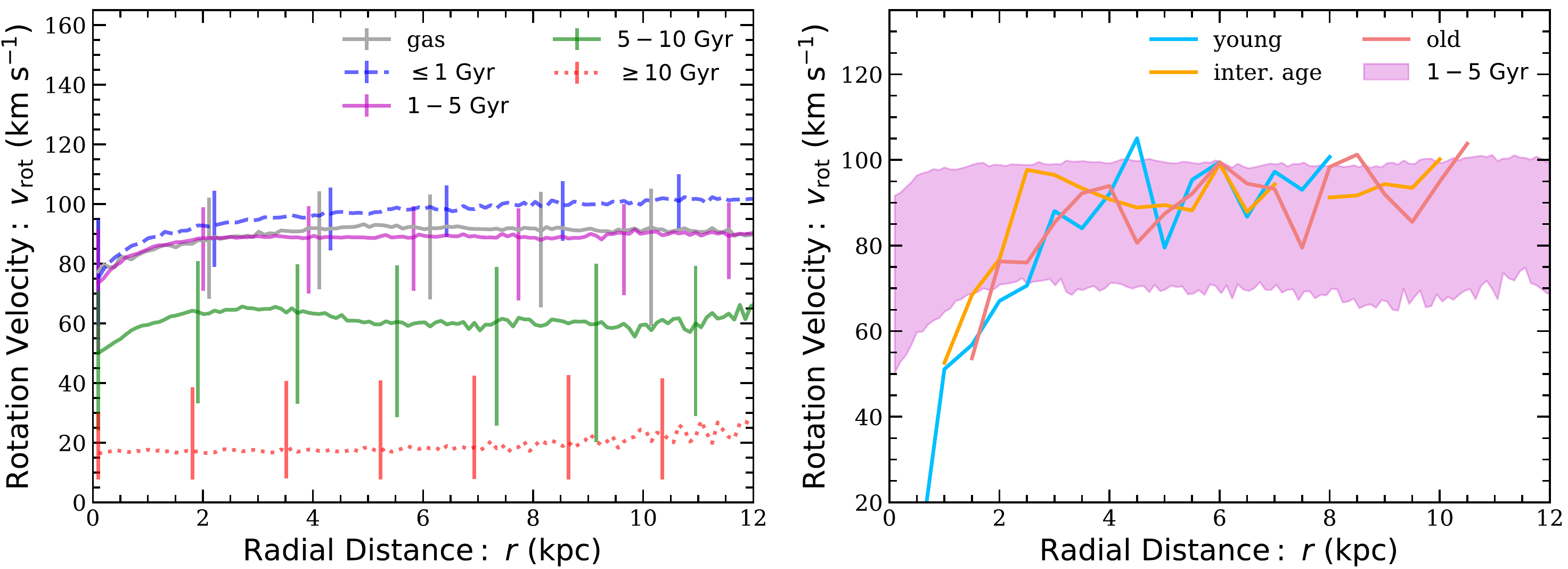}
    \caption{Cumulative stellar rotation curves for the 216 M33-like analogs from the IllustrisTNG50-1 simulation and the observations presented in this analysis. Left panel: The grey line represents the gas cells, the blue dashed line represent star particles with ages $<1$ Gyr, the purple line represents star particles with ages 1-5 Gyr, the green line represents star particles with ages 5-10 Gyr, and the red dotted line represents star particles with ages $>10$ Gyr. The vertical bars represent the width of the distribution ($16^{\rm th} - 84^{\rm th}$ percentile) of rotation velocities in a given radial bin. The median rotation velocities of the 0.1 kpc radial bins are shown. Right panel: The magenta filled region represents stellar particles with ages 1-5 Gyr. The width of the filled regions represent the width of the distribution ($16^{\rm th} - 84^{\rm th}$ percentile) of rotation velocities in a given radial bin. The observed M33 rotation curves for the three age bins are plotted on top. The observations are consistent with stellar particles ages 1-5 Gyr, which are a similar age to the observed sample.}
    \label{fig:TNG_RC}
\end{figure*}

\par For each analog, we rotate the particles/gas cells to be face-on so that the ``line-of-sight'' direction is the $z$ component for all analogs. We exclude particles with $|z| > 10$ kpc and with a galactic radius of $> 20$ kpc to target star particles that likely belong to the disk. We then locally average the velocities to mimic the resolution from observations, following Section 3 of \cite{Quirk2020}. We divide the stellar particles into four groups based on exact stellar age: $< 1$ Gyr, 1-5 Gyr, 5-10 Gyr, and $> 10$ Gyr. Unlike the rough age division used for the observational part of the analysis, these age bins use a star particle's exact age and spans the full cosmological time. We use the gas cells to calculate AD for each age bin as well. 

\par For each analog, we use the smoothed kinematics to construct an azimuthally averaged rotation curve using the equations below. 

\begin{equation}
    v_{rad} = \frac{x \cdot v_{x} + y \cdot v_{y}}{\sqrt{x^{2} + y^{2}}}
\end{equation}

\begin{equation}
    v_{tot} = \sqrt{v_{x}^{2} + v_{y}^{2}}
\end{equation}

\begin{equation}
    v_{rot} = v_{tan} = \sqrt{v_{tot}^{2} - v_{rad}^{2}}
\end{equation}

We limit the analysis to the 2D $xy$ plane to mimic the observed line-of-sight velocities. We choose to azimuthally average the rotation curve for each. To azimuthally average the rotation curves, the particles/cells are placed into 0.1 kpc bins based on their distance from the center of the analog. We then take the median of the rotation velocities in each bin for the final rotation curve. This process creates a rotation curve similar to the median lines shown in Figures \ref{fig:rc_full} and \ref{fig:rc_inner}. 

\begin{figure}[bp!]
\centering
    \includegraphics[width=\columnwidth]{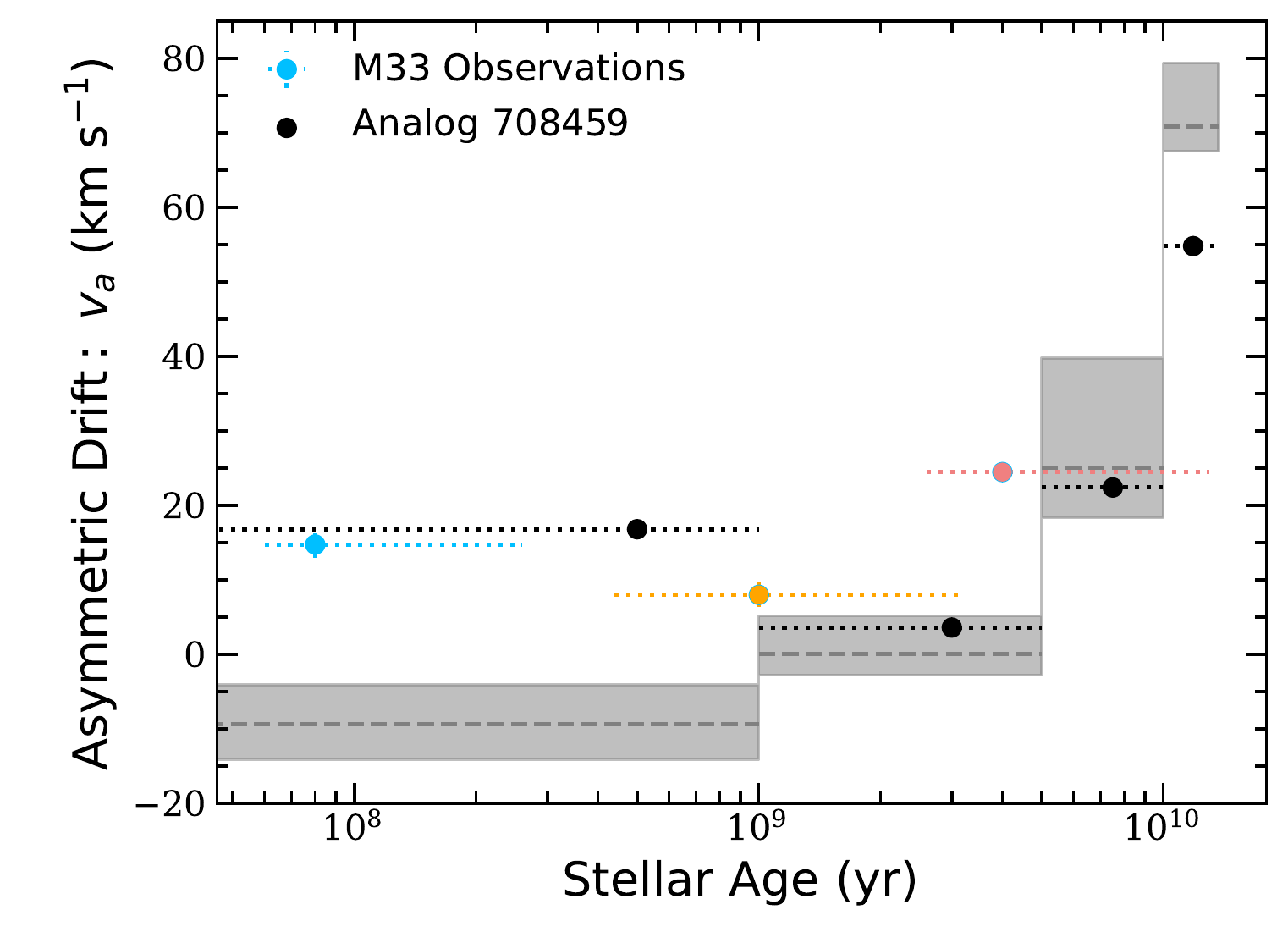}
    \caption{Cumulative median AD measurements for M33-like analogs and for observations of young, intermediate age, and old stars in the disk of M33. The shaded grey regions represent the median AD and the widths of the distribution ($16^{\rm th} - 84^{\rm th}$ percentile) of the AD for the analogs. The color points represent the median AD for the young (blue), intermediate age (orange), and old (red) stars. The black points represent the analog with AD values closest to what is observed. Compared to the whole sample of analogs, the AD from observations is significantly higher for the youngest stars, than that is seen in the simulated analogs.}
    \label{fig:TNG_comp}
\end{figure}

\par To calculate a single set of rotation curves and median AD values as a function of stellar age for the entire analog sample, we first calculate both the rotation curve and AD for each individual analog, as described above. Since the rotation curves are radially binned, the cumulative rotation curve for the entire sample of analogs is made by using the median rotation velocity for each stellar particle age group at each radial bin. We calculate AD values for the entire sample using this cumulative rotation curve. The set of cumulative rotation curves is shown in the left panel of Figure \ref{fig:TNG_RC}. Also shown in this figure are the median rotation curves for the observed M33 young stars, intermediate age stars, and old stars in the right panel. 

\par The set of observed M33 rotation curves is consistent with the rotation curves for star particles with ages $1-5$ Gyr from the simulated analogs. However, these rotation curves are not a perfect comparison, as the observational one comes from a tilted ring model, and the simulated ones come from calculating the actual tangential velocity of the stellar particles.

\par The cumulative AD values for the simulated M33 analogs as a function of stellar particle age are shown in Figure \ref{fig:TNG_comp} along with the observational median M33 AD measurements. For the simulated analogs, AD increases with stellar age. A gradient in AD with stellar age is also seen in M31-like simulated analogs in the same simulation \citep{Quirk2020}. There is no such gradient in the observed AD in M33.

\par For the young and intermediate age stars, the observed AD is higher than what is seen in simulated M33-like analogs. The AD for the oldest stars is consistent within the observed age errors. While some of the differences in the observed M33 AD and the simulated analogs AD is most likely from the differences in the way AD is calculated for each, it is likely that this would affect all of the age bins equally. Thus, the difference in AD between the observed intermediate and the simulated analogs could be a nonphysical systematic offset. However, the observed young stars have an AD value that is more offset from the AD of simulated analogs than the offset for the other two age groups. Thus even if there is a systematic offset, the young observed stars still break the gradient in age and AD that is seen in the simulated M33-like analogs. 

\par The high AD observed in the young stars in M33 suggests that some phenomenon is dynamically heating the young stars. Of the M33-like analogs, seven have young stars with an AD greater than 10 \kms. Six have young star AD values between 12 and 19 \kms, and one has an AD value of 33 \kms for the young stars. Of these seven, four have AD distributions that are similar for all stars 0-10 Gyr (the age bins that best represent the observed bins). For three analogs, the intermediate age stars  (1-5 Gyr) have the lowest AD. Upon visual inspection of the star and gas particles used in these analogs, only one analog shows a slight disturbance in the disk, and only one has had any mergers (minor or major) in the past 4 Gyr. Thus, there is no immediate connection between these analogs that could point towards a cause of the high AD in the young stars. Figure \ref{fig:TNG_comp} shows the AD values for the analog that is closest to the observed AD values. This analog has not had any mergers within the past 4 Gyr and does not show a disturbance or asymmetry in its disk. Closer inspection of the stellar assembly history of this analog is future work.

\par The observed sample of stars and the sample of simulated star particles are meant to exclude possible halo stars. However, halo cuts for each are different, and those for the simulated star particles were not based on a kinematic model. Including all halo stars roughly doubles the AD value for the intermediate age and old bins (see Section \ref{sec:LG} and Table \ref{tab:AD_ill_comp}). It is possible that the gap between the AD in the observations and the simulated M33-like analogs is larger for the intermediate age and old stars, but even if the simulated AD halved for the older bins, the largest gap would still be for the young stars. 

\section{Discussion: Contextualizing M33 in the Local Group}
\label{sec:LG}
\par Resolved stellar spectroscopy of individual stars allows us to study a galaxy in detail. For example, one can examine dynamics as a function of age, spatial position, and metallicity, which puts narrow constraints on origin and evolution. Studies of this kind have already proved valuable for the comparative evolution of the LG's two massive spirals, M31 and the MW. In spite of having similar masses ($M_{\rm vir} \approx\ 10^{12}\ M_{\odot}$), the disk kinematics of the MW and M31 differ significantly. The MW has thin disk stars, while the majority of disk stars in M31 belong to a thick disk or kicked up disk \citep{Dorman2013, Dalcanton2015}, with all the analysis pointing towards M31 having a more violent merger history than the MW \citep[e.g.][]{Tanaka2009, Mackey2019, Ferguson2016}. 

\par \citet{Leaman2017} use data from the literature to compare the age-velocity dispersion relation of eight LG members, including the MW, M31, and M33 to model the evolving interstellar medium (ISM) turbulence as a galaxy experiences heating from various sources. In this section, we focus on the observed velocity dispersion of the three most massive LG members and add measurements for M33 that include a robust sample of individually resolved stars across the inner and outer disk of M33. We also present the first comparison of AD as a function of stellar age of LG members.

\begin{figure}
    \centering
    \includegraphics[width=\columnwidth]{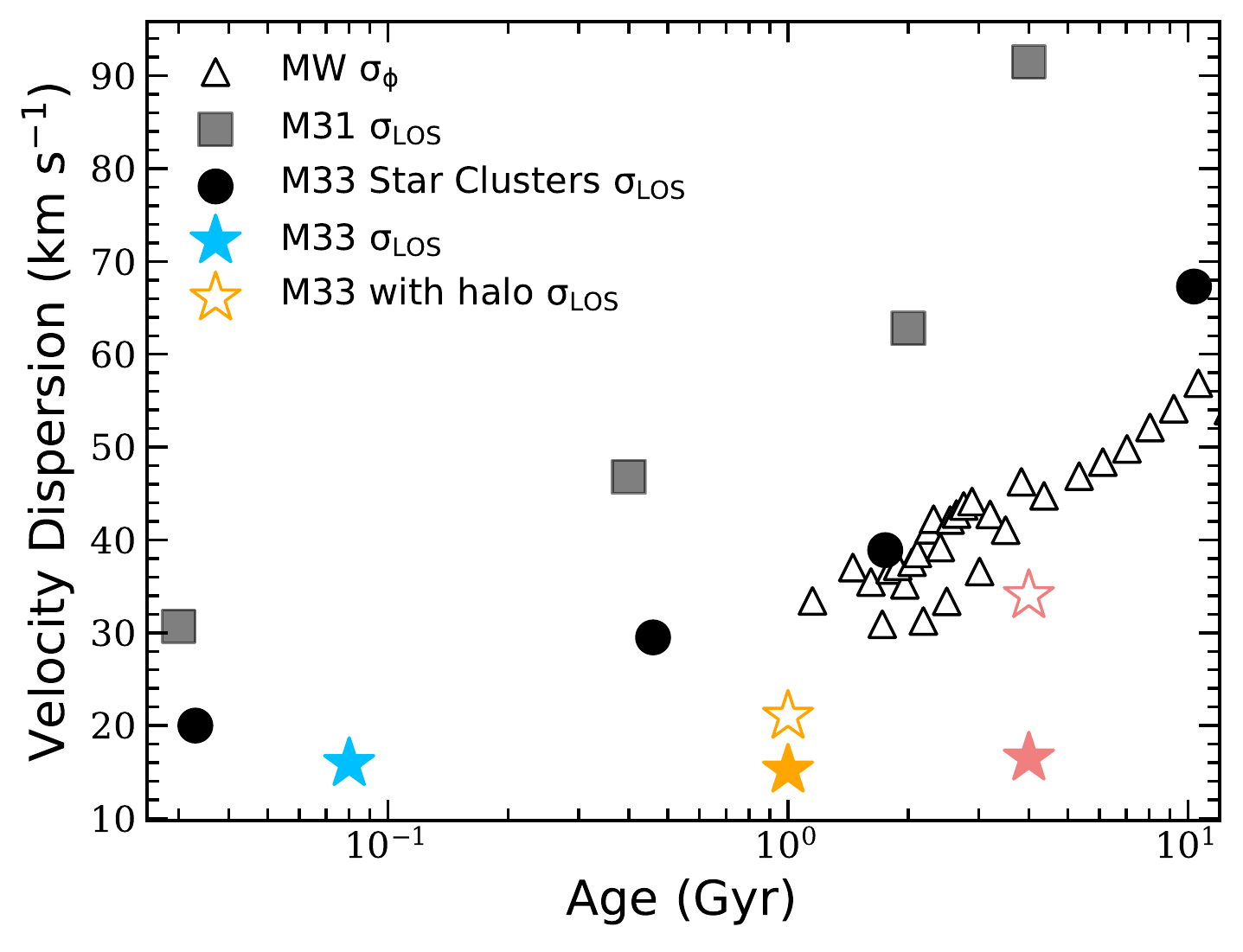}
    \caption{A comparison of velocity dispersion as a function of stellar age for the MW \citep[black triangles]{Nordstrom2004}, M31 \citep[black squares]{Dorman2015}, M33 star clusters \citep[open circles; data later used by \citet{Leaman2017}]{beasley2015}, and M33 (this work, stars). The open stars show the velocity dispersion for intermediate age and old stars in M33 if halo candidates are not removed. For M31 and M33, the velocity dispersion is in the line-of-sight, whereas for the MW, the azimuthal component of velocity dispersion is shown. The MW measurements are also in the solar neighborhood, while the measurements of M31 come from across the northern disk, and those from M33 come from across the entire disk. The MW and M31 show a clear increase in velocity dispersion with stellar age. M33 star clusters also show this increase, but the data from this work does not. Additionally, the magnitude of velocity dispersion from this work is much lower than that seen in the M33 star clusters.}
    \label{fig:LG_disp_comp}
\end{figure}

\par It is important to note that possible halo stars were not removed from the M31 velocity dispersion or AD analysis. 
\citet{Dorman2012} find that $\sim 15\%$ of stars in the M31 sample are part of a dynamically hot component. \cite{Dorman2013} characterize the bulge, disk, and halo of M31 using the luminosity function of stars from the PHAT survey, line-of-sight velocities of RGB stars from the SPLASH survey, and surface brightness profiles. Using these criteria, \citet{Dorman2013} find that the hot component in the M31 sample is best described by a disk-like surface brightness profile and luminosity function but spheroid-like kinematics, suggesting that $\sim 30\%$ of the dynamically hot stars belong to a kicked up disk and $\sim 10\%$ of the M31 sample is halo contamination. Since \citet{Dorman2013} did not fit for a both a disk and a kicked up disk along with a halo component, it is possible that the halo contamination is lower.

\par The MW data presented here consists of nearby F and G dwarf stars in the Solar Neighborhood that were targeted using Hipparcos parallaxes as part of the Geneva-Copenhagen Survey \citep{Nordstrom2004}. \citet{Nordstrom2004} use kinematics to identify thick disk stars and distances to target nearby stars but do not formally remove possible halo stars. The MW halo fraction in the Solar Neighborhood is very small \citep[$0.15\%$][]{du2018}, so it is possible the sample contains few halo stars but unlikely that these bias the velocity dispersion measurements significantly.
\par It was important for this work to remove possible M33 halo stars because M33's halo is centrally concentrated and dense, so there is more overlap between the halo and disk stars' sightlines, giving a larger halo fraction in the RGB stars in M33. For the sake of comparison, though, we show the median velocity dispersion and AD values for M33 if halo stars are not removed. We also acknowledge there is not necessarily a clear distinction between a galaxy's halo and kicked up disk, even if they were formed from different mechanisms and that differentiating between a disk, kicked up disk, and a halo can be arbitrary. 

\par Figure \ref{fig:LG_disp_comp} shows velocity dispersion as a function of stellar age for the MW solar neighborhood \citep{Nordstrom2004, Holmberg2009}, the northern disk of M31 \citep{Dorman2015}, star clusters in M33 \citep{beasley2015}, and stars in the disk of M33 (this work). M31 has both a greater velocity dispersion \citep{Nordstrom2004, Holmberg2009, Sysoliatina2018, Dorman2015, Budanova2017} and a steeper gradient in the age-velocity dispersion than the MW and M33 \citep{Dorman2015, Bhattacharya_2019}. The median velocity dispersion values are significantly lower in M33, which is expected, as M33 is a less massive and therefore more fragile disk. 

\begin{table*}[]
    \centering
    \begin{tabular}{c|c|c|c}
    \hline
    Age Group & AD w.r.t \hi\ in & AD in M33 & AD w.r.t \hi\ in\\
    & M33 ($\rm kms^{-1}$) & inc. halo ($\rm kms^{-1}$)& M31 ($\rm kms^{-1}$) \\
    \hline 
        Young & $15.1_{-1.8}^{+1.6}$ & $15.1_{-1.8}^{+1.6}$ & $-8.2_{-0.72}^{+0.74}$ \\
        Intermediate Age & $8.0_{-1.6}^{+1.7}$ & $20.6_{-1.8}^{+1.9}$ & $34.1_{-4.0}^{+4.4}$ \\
        Old & $24.5_{-1.1}^{+0.9}$ & $42.0_{-1.2}^{+1.1}$ & $63.0_{-0.4}^{+0.59}$ \\
    \hline
    \end{tabular}
    \caption{Median values of AD for stars in M33 (this work) and in M31 \citep{Quirk2019}.}
    \label{tab:LG_comp_ad}
\end{table*}\label{tab:AD_ill_comp}

\par The velocity dispersion reported in this work is not consistent with that measured using star clusters in M33 \citep{beasley2015}. The star clusters show a higher magnitude that is consistent with the MW Solar Neighborhood and a gradient in velocity dispersion and stellar age. One potential explanation for the difference is that we remove stars with line-of-sight velocities that are consistent with a dynamically hot component. If we do not remove these stars, we do recover a sloped age velocity dispersion relation and higher magnitudes, although not high enough to match the dispersion of the star clusters. It is possible that a larger fraction of the star clusters than individual stars belong to the hot component. 

\par It is unexpected that this work does not see a trend between velocity dispersion and age, as it is predicted and seen in the MW and M31 \citep{Leaman2017, beasley2015, Nordstrom2004, Dorman2015}. This lack of trend is what \citet{Leaman2017} calculate for dwarf spheroids in the LG with masses $\sim 1000$ times less than M33, not for a more massive galaxy like M33. Their models, which do not distinguish between galactic components, predict M33 should have a velocity dispersion of $\sim 40$ \kms for stars with ages of several Gyr, which is also consistent with the observations of M33 star clusters \citep{beasley2015}. If we do not remove likely halo stars, the oldest M33 stars in this work have a velocity dispersion that approaches 40 \kms. This demonstrates the effect that contaminate halo stars can have in a measured age-velocity dispersion relation. The velocity dispersion of the individual stars and the clusters were also measured in two different ways in two different samples, which results in different systematic uncertainties in the intrinsic dispersion calculation and could contribute to the discrepancy.

\par The existence of a substantial kinematically hot halo component in M33's inner disk \citep{Gilbert2021} may suggest internal heating mechanisms, like gas flows from stellar feedback \citep{el-badry2016}, could have caused enough disk heating to substantially contribute to a stellar halo, which could also cause stellar clusters and individual stars that were born in the disk to be heated or kicked up into the halo component. 

\par Additionally, because of their respective inclinations, the observed line-of-sight velocity dispersion of M33 includes a larger $z$ component than the observed line-of-sight velocity dispersion of M31. \citet{martig2014} find that in more massive disks, the vertical component of velocity dispersion is largely dependent on merger history but these trends can be obscured when there are large uncertainties on age estimates. It is possible our wide age bins combined with the large influence of the vertical component in M33's line-of-sight velocity dispersion are obscuring a subtle trend in the age-velocity dispersion relation, but it is unlikely to fully explain the low magnitude of velocity dispersion for the intermediate age and old stars.

\par We also make the first comparisons of the AD as a function of stellar age observed in M31 and M33. There are no current measurements of AD as a function of stellar age in the MW or other galaxies in the LG. Table \ref{tab:LG_comp_ad} compares the AD (with respect to \hi) in M31 and M33. Like velocity dispersion, M31 has greater values of AD and a steeper rise in AD as a function of age than observed in M33, suggesting that it has experienced several significant and ongoing heating events throughout its relatively recent history. If possible halo stars are not removed from the M33 dataset, there is a trend between velocity dispersion/AD and stellar age. However, this trend disappears if the likely halo stars are removed. Furthermore, the AD values roughly double when possible halo stars are not removed for the intermediate age and old stars. 

\par There are other measurements of AD from Integral-field-units (IFUs): \cite{Martinsson2013} use IFUs to measure the AD with respect to ionized gas in face-on spiral galaxies and find stars lag behind the ionized gas by $\sim 11\pm8\%$, which is consistent with the young stars lagging on average by $20\%$, the intermediate age stars by $9\%$, and the old stars by $18\%$. This amount of lag is similar to studies of AD in other inclined local galaxies \citep{Ciardullo2004,Herrmann2009,Westfall2007, Westfall2011} and in the MW \citep{Ratnatunga1997,Olling}.

\section{Summary and Conclusions}
\label{sec:summary}
In this work, we present the largest stellar spectroscopic survey of M33, the TREX Survey, and present initial analysis of the stellar disk kinematics as a function of stellar age using only individually resolved stars, which is the first of its kind in M33. Below we summarize our main findings and conclusions of the complex, yet low mass, galaxy.
\begin{itemize}
\item 1. The TREX survey consists of $\sim 4500$ stars with good quality spectra across the entire disk of M33, ranging from several evolutionary stages: massive MS and HeB stars, intermediate mass AGB stars, and low mass RGB stars. This work uses a subset (2561 spectra) of the full survey.
\item 2. We find that M33's stellar disk has an average velocity dispersion of $\sim 16$ \kms, which is significantly lower than what is observed in the disk of MW and M31 \citep{Holmberg2009, Dorman2015} and lower than what is measured using star clusters \citep{beasley2015}. The average magnitude of AD is on the order of $\sim 16$, which is also lower than what is observed in M31. 
\item 3. Velocity dispersion and AD do not increase with stellar age in the disk, which is unexpected. We highlight the importance of removing potential halo stars when measuring the age-velocity dispersion and age-AD relation. 
\item 4. This analysis suggests that M33 has experienced a significantly different dynamical heating history than M31 and the MW, which may have been dominated by internal heating mechanisms rather than external ones. 
\item 5. The young stars are as dynamically hot as the older stars in the stellar disk component. These young stars also have a wider distribution of AD values than the old disk stars. They are more dynamically hot than simulated M33-like analogs predict. Possible mechanisms for the heating of the young stars could include turbulence from ongoing star formation or scattering from giant molecular clouds. It is also possible that these stars are remnants of a relatively recent minor merger or other type of galaxy interaction and lie in front of the disk.
\end{itemize}

\section*{Acknowledgements} 
The authors would like to thank everyone who contributed to the general public's safety, health, and well being during the ongoing COVID-19 pandemic. A very large thank you to all of the essential workers and medical professionals. Thank you also to the folks at Keck who worked to make PJ observing happen. 

The authors recognize and acknowledge the very significant cultural role and reverence that the summit of Maunakea has always had within the indigenous Hawaiian community. We are most fortunate to have the opportunity to conduct observations from this mountain.

Thank you to Bruno Villasenor, without whom, our code would have been too slow. ACNQ and PG are grateful to the Sulphurous AA Overspray project for their generous support.

We would like to thank the anonymous referee for improving the clarity of this paper.

Support for this work was provided by NSF grants AST-1909066 (KMG), AST-1909759 (PG), DGE-1842400 (ACNQ),  GO-14610 (BW), and ANID/FONDECYT Regular Project 1210992 (LC). The analysis pipeline used to reduce the DEIMOS data was developed at UC Berkeley with support from NSF grant AST-0071048. \\

\noindent {\it Facilities:} Keck:II(DEIMOS), CFHT(MegaCam), HST(ACS)\\

\noindent {\it Software:} This research made use of \texttt{astropy} \citep{astropy2013, astropy2018}, \texttt{matplotlib} \citep{Hunter:2007}, \texttt{numpy} \citep{numpy}, \texttt{scipy} \citep{scipy}, \texttt{scikit-learn} \citep{scikit-learn}, and \texttt{Illustris Python} \citep{nelson15}. \\

\noindent {\it Data Availability Statement:} The data used in this article is from the PAndAS catalogue, which is available to download at: http://www.cadc-ccda.hia-iha.nrc-cnrc.gc.ca/en/community/pandas/query.html. The other data in this work come from sources available on the Keck and HST archive or are not publicly available. \\


\bibliography{M33_Dynamics}{}
\bibliographystyle{aasjournal}

\appendix 
\section{Special stars: weak CN and carbon Stars}
\label{sec:rare}
As mentioned in Section \ref{sec:ages} we identified two classes of rare M33 stars with unusual spectroscopic features, carbon stars and ``weak CN'' stars in our TREX survey dataset, using the same combination of visual inspection and machine classification of spectra that our team has used to identify such stars in M31 \citep{Hamren_2015, Hamren_2016, Guhathakurta2017, Kamath2017, Masegian2019, Kotha2020, Girardi_2020, Cbauhan2021, Rodriguez2021}.
The carbon stars are identified by their strong CN, CH, and C$_2$ molecular absorption bands, the strongest of which is the ``W''-shaped CN band at 8000\ \AA. By contrast, the ``weak CN'' spectrum displays a much weaker 8000\ \AA\ CN absorption feature while the rest of the spectrum resembles that of a normal (i.e., O-rich) star of comparable color in that it displays the near-infrared Ca triplet and, for redder stars, a set of TiO bands.

\par We have compared the positions of these rare M33 stars in CMDs to theoretical stellar tracks from the PARSEC library \citep{Bressan1993, Bressan2012}, for a variety of combinations of the PHATTER filters (all except F275W). As expected, this comparison indicates that carbon stars are intermediate mass/intermediate age AGB stars with masses of $M \sim 2$--4 $M_\odot$ and ages of a few Gyr. In contrast, the ``weak CN'' stars are massive ($M \sim 4$--$M_\odot$) and short lived ($t<\rm 100$ Myr) in the RGB/AGB evolutionary phase. 

\par We identify 97 carbon stars and 363 weak CN stars that pass our quality and disk star cuts. For the bulk of this analysis, we have assigned the weak CN stars to the young star group and the carbon stars to the intermediate age group. In this section, we examine the kinematics of these carbon and ``weak CN'' stars separately to see if they are unique beyond their spectral features. 

\par
\begin{figure}[bp!]
    \centering
    \includegraphics[scale=.6]{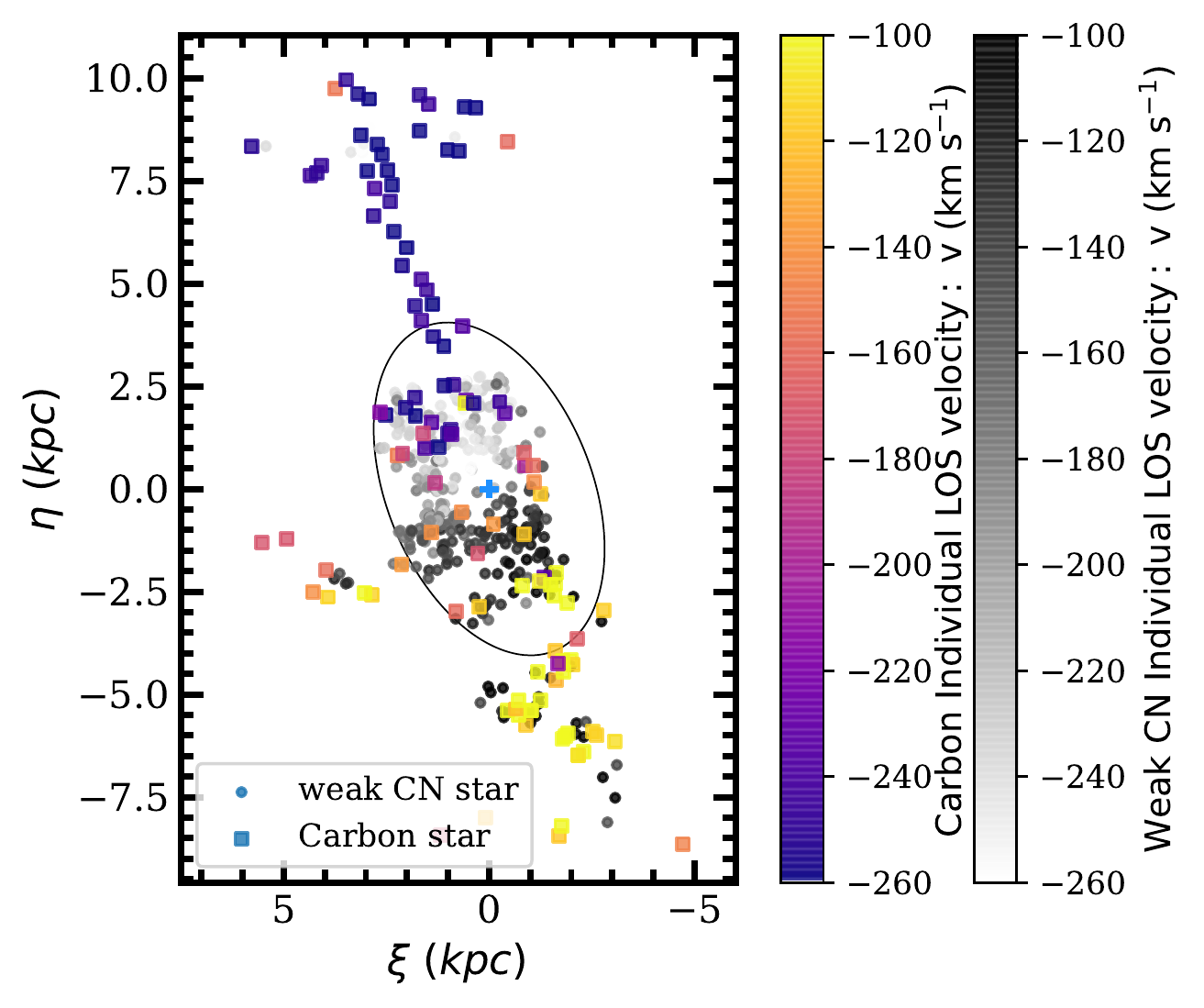}
    \caption{A spatial map showing the location of identified weak CN and carbon stars. The color represents the line-of-sight velocity. The weak CN stars have circular symbols and a grayscale colorbar, and the carbon stars have square symbols and a color colorbar.}
    \label{fig:special_stars_map}
\end{figure}

 Figure \ref{fig:special_stars_map} shows the location and line-of-sight velocity of the weak CN and carbon stars. As there is a lack of young stars in the outer disk in our sample due to different target selection epochs, the weak CN stars are centrally concentrated while the carbon stars are present at extended radii. We have calculated the velocity dispersion for these two groups (Table \ref{tab:special_stars_ad}). The weak CN stars have a lower velocity dispersion (11.9\kms) than the young stars, intermediate stars, and old stars (15.9 \kms, 15.2\kms, and 16.5\kms, respectively; Table \ref{tab:LOS_disp}), and the carbon stars have a slightly higher velocity dispersion (20.3\kms) than all the three age groups. 

\par We also construct rotation curves and calculate AD for the weak CN and carbon stars. Because most of the carbon stars are located at large radii, we only calculate AD with respect to the \hi. The median AD values, width of the distributions, and percentage of outliers in the distribution are shown in Table \ref{tab:special_stars_ad}. The AD of the weak CN stars (8.0\kms) most closely resembles that of the intermediate age stars (8.0\kms). The carbon stars' median AD value (22.4\kms) is more consistent with the old (24.5\kms) than the intermediate age stars, even though they are believed to be intermediate age. The width of the distribution of AD for the weak CN and carbon stars are both closest to the width of the intermediate age stars (67.7\kms), which is wider than the old stars. There does not appear to be anything kinematically unique about these rare spectral types.

\begin{table}[]
    \centering
    \begin{tabular}{c|c|c|c|c}
    \hline
    Rare Star & $\sigma_{LOS}$ & AD w.r.t \hi\ & Width & Outliers\\
    & ($\rm kms^{-1}$)  & ($\rm kms^{-1}$) & ($\rm kms^{-1}$) & (\%)\\
    \hline 
        weak CN & $11.9_{-0.3}^{+0.4}$ & $8.0_{-2.6}^{+2.2}$ & 70.0 & 3.5 \\
        carbon & $20.3_{-1.5}^{+1.2}$ & $22.4_{-3.3}^{+23.5}$ & 58.8 & 0.0 \\
    \hline
    \end{tabular}
    \caption{Median values of velocity dispersion and AD for the weak CN and carbon stars with respect to \hi. Also shown is the width (sigma) of the AD distributions and the percentage of outliers in the distributions. The errors on the median value represent the difference between the $16^{\rm th} \rm and\ 84^{\rm th}$ percentiles divided by $\sqrt{\rm N}$, where N is the number of stars.}
    \label{tab:special_stars_ad}
\end{table}



\end{document}